# Fractals in the Nervous System: conceptual implications for Theoretical Neuroscience.

Gerhard Werner
gwer1@mail.utexas.edu
Department of Biomedical Engineering
University of Texas at Austin, TX.

Abstract

This essay is presented with two principal objectives in mind: first, to document the prevalence of fractals at all levels of the nervous system, giving credence to the notion of their functional relevance; and second, to draw attention to the as yet still unresolved issues of the detailed relationships among power law scaling, self-similarity, and self-organized criticality. As regards criticality, I will document that it has become a pivotal reference point in Neurodynamics. Furthermore, I will emphasize the not yet fully appreciated significance of allometric control processes. For dynamic fractals, I will assemble reasons for attributing to them the capacity to adapt task execution to contextual changes across a range of scales. The final Section consists of general reflections on the implications of the reviewed data, and identifies what appear to be issues of fundamental importance for future research in the rapidly evolving topic of this review.

Contents

## 1. Introduction

Fractals, introduced by Mandelbrot in 1977, are in the spatial domain considered to be self-similar geometric objects with features on an infinite number of scales. In the analysis of time series, fractal time describes highly intermittent self-similar temporal behavior that does not possess a characteristic time scale. Their statistical analysis can provide access to understanding the dynamics of complex systems. Not possessing a single characteristic scale, static and dynamical fractals, measured on different scales of space and time, respectively, can be characterized by power functions whose (usually non-integer) exponents are their fractal Dimensions. In this essay, principal emphasis is on random fractals which include a stochastic element in their generation.

Fractals are signals which display scale-invariant, self-similar behavior; they can be analyzed by decomposing a signal into a hierarchy of temporal and spatial scales that may cover the wide range between coarse-scale long-term and high-frequency fine-scale fluctuations. If the relationship of the property under consideration is simple with respect to change of scale, the process is considered monofractal , the process can then be characterized by a single scaling exponent which is related to the fractal dimension or the spectral exponent of the process. It can be expressed as the Hurst exponent of the process (Mandelbrot, 1968; Koutsoyiannis, 2002). However, in some instances the scaling behavior may not be adequately characterized by a single, stationary scaling exponent. In such cases, several scaling exponents may be required, each exponent locally pertaining to some portion of the data stream. Such multifractal signals are represented by the histogram of the Hoelder exponents, also known as the singularity spectrum (Muzy et al., 1993).

Power law scaling and other manifestations of fractal and self-similar patterns in space and/or time can be identified at all levels of neural organization. With few exceptions, these observations remained largely islands in the otherwise rapidly advancing theoretical Neuroscience with different priorities. However, recent advances in methodology of measurement of fractal connectivity at higher levels of brain organization have led to a proliferation of new data. This now calls for integrating fractality with other insights into brain organization and complexity, notably in the light of the substantial evidence for the brain being a complex system in a regime of criticality, as understood in statistical physics (Chialvo,2004, 2008; Kitzbichler et al, 2009; Fraiman et al, 2009; Werner, 2007b, 2009a,b). Like in other physiological systems manifesting fractal patterns (see for instance: Bassigthwaighte et al, 1994; West and Deering, 1995; Iannacone and Khoka, 1995; West, 2006) the question of ubiquity of power law scaling needs to be addressed in relation to other features of brain organization. Similarly, is there a relation between fractal organization and the propensity for phase transitions of critical systems? Is there a bridge between coarse graining (including renormalization group transformation) and fractality ? And, most importantly, can fractal properties be viewed as playing a role for the functional integration among different levels of neuronal organization as Andersen (2000) suggests in an article entitled "From Molecules to Mindfulness". Generalizing from a comprehensive theory of organization and interactions at the molecular level, Agnati et al (2004, 2009) view the Central Nervous System as a nested network at all levels of organization, in the image of the Russian Matryoshka dolls: self-similar structures being embedded within one another. This theory is elaborated in great detail, with a "Fractal (self-similarity) Logic" operating on a set of identical rules which would govern the relation between successive levels of the nested system.

Giesinger's ( 2001 ) comprehensive overview of scale invariance in Biology provides the background of this review, as do the insights gained in Physics through the work of Wilson (1979 ) and Kadanoff (1990), amongst many others. In addition, Turcotte (1999) discussed at great length the relation between aspects of self-organized criticality and fractal scaling from the points of view of "avalanche" behavior and systematic properties of correlation length, specifically directing attention to inverse cascade and site-percolation models of the well-known forest fire paradigm. While none of the issues discussed in the following will receive a definitive answer, I will aim at an explicit formulation of the network of interrelated factors that constitute the territory in which new perspectives and potential solutions may lie.

With the agenda set forth in the foregoing, the organization of the presentation is as follows: I will first briefly review the neuroscience literature on fractals, organized by level of neuronal organization, from ion channels to cortical networks, and to psychological and behavioral functions. Criticality of brain states, Allometric control and the origin of fractals by phase transitions of complex systems will be addressed in more detail. This will be followed by a brief overview of the essentials features of the theory of fractal generators, including random walk theory and fractional differential operators. Having laid out the background in this manner, I will consider relations between renormalization group transformation and fractals as having some potential bearing on the apparent ubiquity and universality of power law scaling in neural structures and processes, and its relation to criticality. Finally, I will direct attention to the amazing consequence of self-similarity which assures the telescoping of different levels of structural and functional organization to constitute a fractal object or time series. This will lead me to posing the ultimate question: is there a process for unpacking interactions between different levels of the fractal object, responsive to circumstances and conditions, which eludes us entirely? If it existed, fractals would surely be a most extraordinary design principle for operational economy in complex systems.

## 2. Power-law scaling in neuronal structures and processes.

This section is intended to summarize essential aspects of fractal properties at each of the conventionally designated organizational levels, as the basis for conceptual consideration of relations across these levels. However, a word of caution is in order: the sketches of observational data in this section encompass a vast variety of biological substrates, conditions of observation, and methods of measurement. This heterogeneity imposes limits on generalizations, as do the differences of criteria for identifying fractal or self-similar features in the data. Potential pitfalls were discussed and illustrated in LaBarbera's (1989) useful (largely pedagogic) publication. More recently, Eke et al (2002), Deligniers et al (2006) and Clauset et al (2009) set forth stringent criteria for design, collection and interpretation of data for identifying and categorizing fractal properties. The latter authors are very specific in formulating a principled statistical framework, combining maximum likelihood fitting methods with goodness of fit tests; they demonstrate examples of data that had been conjectured to represent power law fits, but did not withstand the rigor of their tests. Touboul and Destexhe (2009) also suggest that apparent power law scaling may in some instances not be supported by more stringent statistical criteria. Of particular relevance to the topic of Section 2.3 is their claim that experimentally observed power law scaling must not considered proof of self-organized criticality, lest there be other supporting evidence available. The analysis, synthesis and estimation of fractal-rate stochastic point processes is reviewed and illustrated with examples by Thurner et al. (1997). Note also that the distinction between mono- and multifractal scaling is sometimes difficult to draw (Kadanoff et al, 1989).

Conceivably, some of the variations among the reports reviewed in subsequent sections may be attributable to procedural differences among studies; other reports may not meet the rigorous statistical criteria of Clauset et al. Nevertheless, I submit that the majority of experimental data on fractals in neural structures give collectively adequate reason for ascribing to them wide-spread functional significance. At least the results based on wavelet analysis appear immune to methodological criticism (see Section 2.4).

### 2.1 Neuronal morphology

In the foundational work “The fractal geometry of Nature”, Mandelbrot (1977) wrote “it would be nice if neurons - he mentioned specifically Purkinje cells in the cerebellum- turned out to be fractal”: Nature obliged abundantly as the following sample of findings with dendrites, neuron cell bodies and glia cells indicates. Studying the branching pattern of dendritic trees of retina neurons, Caserta et al (1990) identify by box counting fractal shapes with a fractal dimension of approximately 1.7, which can be explained by a diffusion limited aggregation model (Witten and Sander, 1981); but fractal dimension measured by different methods (for instance comparing box counting with cumulative mass method) gives appreciably different values (Caserta et al, 1955). A fractal structure was observed by Kniffki et al (1994 ) for the branching dendrite patterns of thalamic neurons in Golgi impregnated specimens. In a separate series, a scaling relation for bifurcations within the dendrite trees was ascertained (Kniffki et al, 1993). Significant species differences in fractal dimensions of dendrite arborizations in dorsal horn spinal cord neurons (Milosevic et al, 2007) may be attributable to species differences in peripheral somesthetic sensibility (the dorsal horn neurons being the first receiving station of this type of afferent input). Fractal analysis also reveals a distinct differentiation of neuron types in the different laminae of the dorsal horn (Milosevic et al, 2005). Differences in regional connectivity and functional capacity amongst different regions in visual cortex pyramidal neurons are also associated with marked variation in the fractal dendrite branching structure (Zietsch and Elston, 2005). Fractal analyses provide a measure of space filling of dendrite arbors which, in a study by Jelinek and Elston (2001), differentiates in the macaque visual cortex the two known processing streams between primary and secondary visual area by differences in fractal properties. These investigators had undertaken a meticulous examination of criteria for 'quality control' in studies of this nature, from the stage of pre-processing of tissue specimens to comparative evaluation of methods for determining fractal dimension (Jelinek et al, 1995).Examining the connectivity repertoire of basal dendrite arbors of pyramidal neurons, Wen et al (2009) determined a universal power law scaling for dendrite length and radius, suggesting that the dendrite arbors are constructed by statistically similar processes; moreover, fragments of an arbor are statistically similar to the entire arbor, thus displaying self-similarity. These design features are thought to maximize functionality for a fixed dendrite cost.

Additional evidence comes from digital image analysis which enabled Smith et al (1989) to determine the fractal dimension of neuron contours. Results obtained with conventional methods of scaling analysis are corroborated by Wavelet Packet fractal analysis (Jones and Jelinek, 1989). Multifractals were identified for cortical pyramidal cells while, in comparison, neurons of synRas transgenic mice display less complex arborization patterns (Schierwagen, 2008.). Shape complexity of neurons and elements of microglia in human brain can be classified over a range for fractal dimensions which is different for normal and pathological brains (Karperien et al, 2008 ). The sequence of developmental stages of oligodendrocytes, tracked the basis of their immunoreactivity, parallels changes in fractal dimension (Bernard et al, 2001). Fractal analysis of cell ramification and space filling patterns differentiate microglia cells into two categories, depending on whether their fractal dimension did, or did not increase after brain injury (Soltys et al, 2001).

The scaling law for the cortical magnification factor in primate visual cortex is an illustrative example with functional significance: as is well known, the part of the visual scene corresponding to the eye center is represented densely at the cortex, becoming progressively sparser towards the periphery. It turns out that the scaling law for the sampling density away from area centralis is a power function which assures locating a peripheral target in the shortest time (Koulakov, 2010). As brain size increases, the cortex thickens only slightly, but the degree of sulcal convolution increases dramatically, indicating that human cortices are not simply scaled versions of one another (Im et al., 2008). Changizi (2003) infers several organizing principles of Neocortex from scaling relations among its components: e.g. diameters of neural structures predispose for efficient transport through neuron arborizations; economical wiring reflects well-connectedness within given volumes of neural tissue. These relations are viewed to represent a universal law for scaling that applies to hierarchical complexity and combinatorial systems, generally (Changizi, 2001 b).

As one among several instances of scaling in the cerebral cortex, Changizi (2001 a) also shows that axon cross sectional area increases in Phylogeny with brain size, presumably compensating increase of conduction distances with conduction velocity (see Section 3.1). A synthesis of comparative neuroanatomy with biophysics leads Harrison et al (2002) to conclude that scaling trends in morphological specializations at the cellular level may constitute functional adaptations. One of the examples in support of their thesis is the role of component scaling for managing the conflicting developmental trends of increasing brain size and surface folding on the one hand, and the requirement for optimizing energy requirements and processing speed, on the other. The role of scaling relations for brain growth becomes evident when its scaling relation is disrupted by preterm birth in a dose-dependent, sexually dimorphic fashion ; it directly parallels the incidence of neuro-developmental impairments in preterm infants (Kapellou et al., 2006).

Taken together, the observations surveyed in the foregoing two paragraphs suggest that fractal dimension of neuronal and glia elements bear some relations to developmental, functional and pathological conditions of neural tissue. This warrants a few conceptual considerations: Bieberich (2002) attaches neural-computational significance to the self-similarity of dendritic branching as a platform for economical information compression and recursive algorithms. On the same self-similarity principle, Pellionisz (1989) envisages a fractal growth model of dendritic arbors by iterated code repetition as process for global construction of fractals (see for instance: Barnsley & Dempko, 1985): the essential underlying theme is to both reduce complexity of generating, and at the same time conserving the full richness of the dendrite arbor. I will expand on this principle in later section of this essay. Among the not yet explored implications of dendrite fractal arborizations are the effect they may induce on the dynamics of processes and critical phenomena in dendrite spines for which they are a supporting platform: In Statistical Physics, such effects obtain when the neighborhood relations among interacting elements (for instance: Ising spins or coupled maps (Cosenza and Kapral, 1992) are themselves provided by a self-similar fractal lattices, such as the Sierpinsky Gasket (Gefen et al 1980), rather than an Euclidean geometric base.

In an extension of fractal analysis to features of complex neural structures, Zhang (2006) determined the Magnetic Resonance image-based fractal dimension of white matter of human brain. This method was shown to accurately quantify white matter structural complexity in three dimensions, and detect age-related degenerative changes. Tractography based on Diffusion Tensor Imaging enabled Katsaloulis and Vergenelakis (2009) to determine fractal dimension, self-similarity and lacunarity of neuron tracts in human brain. The lacunarity analysis is understood to indicate the distribution of fractal neuron tracts of different length scales, as evidence of connections between different neuron ensembles. Another extraordinary technical advance made it possible to determine the fractal properties of receptor density and distribution in human brain, using Positron Emission Tomography (PET) and Single-photon Emission Tomography (SPET) (Kuikka and Tiihonen, 1998).

2.2 The peripheral nervous system: ion channels, point process analysis of activity in peripheral nerves and individual neurons

2.2.1: Ion Channels.

Turning to primarily functional aspects of fractality in neural systems, attention focuses in this section on temporal aspects of ion channel gating and its relation to time series of neuronal discharge patterns. The following collage of data obtained with different experimental conditions as well as modeling studies consistently supports the dominant presence of fractal features in the functional manifestations at the levels under consideration. The kinetics of Ion transport across neuronal membranes occurs, in part, via ion channels . Application of the Patch clamp technique made it possible to follow the time course of channel opening and closing precisely. Typically, the rate of channel opening and closing opening fluctuates, changing at times suddenly from periods of great to periods of slow activity. This pattern served as clue to surmise an underlying fractal process with infinite variance. On this basis Liebowitch et al (1987; 2001) asked how the switching probabilities at one time scale of observation are related to those at another time scale. It turned out that these probabilities (defined as effective kinetic rate) at a given time scale are characterized by fractal scaling, and that effective kinetic rates for different time scales of observation display self-similarity: there are bursts within bursts of openings and closings. The suggestion is that energy barriers in stochastically switching protein conformational states are the underlying mechanism (for a detailed account, see Ch. 8 in Bassingthwhaite et al, 1994). A different version that also accounts for the power law relationship of ion channel gating kinetics assumes that ion channel proteins have a very large number of states, all of similar energy, making the gating process more akin to a diffusion (Millhauser 1988). Recent theoretical modeling defined more precisely the conditions that give rise to the power law distributions in relation to the activation barriers, compatible with the known Physics of proteins (Goychuck and Hanggi, 2002). Roncaglia et al (1993) developed on theoretical grounds a stringent criterion for ascertaining the validity of the fractal theory by evaluating the experimental distribution of channel closing times in terms of the Hurst phenomenon. A few years thereafter, Varanda at al. (2000) delivered the evidence for Ca-activated K channels in the form of long term correlations of open and closed dwell times, expressed as Hurst coefficients of the order of 0.6, which alternative Markovian models failed to satisfy.

Before proceeding to discuss the implication of channel kinetics for the patterning of trains of neuron spikes, a brief remark on the fractal activity at the site of neural impulse transmission at the neuromuscular junction. As is well known from the work of DelCastillo and Katz (1954), the neural transmitter substance acetylcholine is released from the nerve terminal in small packages: the miniature end potentials (MEPP) are considered manifestations of the exocytosis of humoral transmitters. In departure from initial textbook accounts of the MEPP release reflecting a set of homogeneous stationary Bernoulli trials, Perkel and Feldman (1979) categorically reject a purely binomial model of (quantal) transmitter release. For the frog neuromuscular junction, Rothshenker & Rahaminoff (1970) could show that excocytosis can exhibit correlations (memory) extending over periods of seconds, suggesting self-similar characteristics. When sampled over prolonged periods, Lowen et al (1997) collected conclusive data at the neuromuscular junction and synapses in hippocampal tissue culture that frequency and amplitudes of MEPP’s display fractal scaling .Takeda et al (1999) also reported comparable findings for the vertebral neuromuscular junction. The detailed analysis of quantitative features of the recorded data led Lowen at al (1997) to conclude that traditional renewal models of vesicular exocytosis as a memoryless stochastic process are entirely inadequate for representing many of its salient features. Instead, their recommendation is that a new class of models should be considered that relies on fractal-rate stochastic point processes: fractal rate activity represents a kind of memory in that occurrence of an event at a given point in time increases the likelihood of another event to occur at a later point in time, with that likelihood persisting for some time.

2.2.2 Point process analysis in peripheral nerves and neurons

In this section, neuron discharge trains are viewed as mathematical objects, belonging to the class of point processes (Thurner et al, 1997; Lowen and Teich, 2005): events occurring at a point in time or space. Werner and Mountcastle (1963, 1964) determined scaling of neural responses in primary cutaneous afferent nerve fibers with the magnitude of mechanical stimuli applied to receptors. The implications of their findings in Psychophysics will be taken up in Section 3. Adaptation in neural structures serves to extend their dynamic range: the significance of this function is discussed in Section 5.2.

The statistics of action potential trains recorded from single neurons in the cochlear nucleus of anaesthetized cats formed the basis of a mathematical analysis by Gerstein and Mandelbrot (1964 ). The principal result was that a random walk model towards an absorbing and a reflecting barrier can account for a wide range of fractal neuronal activity patterns, assuming no more than the known physiological mechanisms of a threshold for membrane depolarization, and the summation of excitatory and inhibitory post synaptic potentials. Except for a thesis by Johannesma in 1969, it took almost 20 years of hegemony of Poisson and Gaussian distributions until fractal approaches to spike train statistics were resumed: this time by Wise (1981) in a study of spike interval distributions of data that had been recorded primarily by Bloom (1969) in the cerebral cortex, and in respiratory neurons recorded by Smolders and Folgering (1977. Wise found that plots of the spike interval histograms on log-log scales showed negative powers on time with long tails, which he attributed to the neuron membrane potential undergoing a random walk while the firing threshold fluctuates. Re-working some of Wise's data, West and Deering (1994) identified fractal (hyperbolic) spike interval distributions. Taking an entirely different approach to conceptualizing irregular behavior in neuron spike trains led Shahverdian and Apkarian (1998) to discuss self-affinity, powerlaw dependence and computational complexity of spike trains in terms of a multidimensional Cantor space with zero Lebesgue measure as attractor.

The turning point in the history of identifying fractal neuronal firing is associated with the work of Teich and Lowen, beginning in the early 1980s (Ch 22, in McKenna, 1992) with invalidating the then prevalent notion of Poisson point processes. More recently, the shortcoming of Poisson spike interval statistics was also pointed out by Kass and Ventura (2001) and by van Vreeswick (2001) who critizised experimental (Richmond et al, 1990) and theoretical (Ohlshausen and Field, 1998) reports for unwarrantedly assuming either Poisson neurons or rate based neurons with rate independent Gaussian noise; instead he considered a renewal model as biologically more plausible.

Teich and Lowen's essential realization was that determining long-time correlations in spike trains requires sample sizes to be appreciably larger than conventionally used. On this basis, Teich et al (1990) identified the following essential features of the time series of neural spikes recorded from cat auditory nerve fibers and the lateral superior olivary nucleus: discharge rates determined with different averaging times can exhibit self-similarity; the variance-to-mean ratio of spike number increases with sufficiently large counting time in a fractional power law fashion, with the exponent in the power law varying with the stimulus level. With these data in hand, Lowen and Teich (1993) suggested that the fractal action potential patterning in auditory nerve may be related to fractal activity in the ion channels of the sensory organs feeding into the auditory nerve: that is, the hair cells in the cochlea. This idea required to show that ion channel gating and neuronal spiking patterns are indeed causally related. Lowen et al (1999) succeeded with demonstrating this causal dependence in computational models, thus adding for the special case of the cochlear hair cells some credence to their proposal that gating patterns in sensory organ ion channels can affect discharge patterns in the sensory nerve tracts they feed. In an elegant experimental design, Teich (1977) not only ascertained a power function for the activity in retina ganglion cells and neurons in the lateral geniculate body when studied independently, but also succeeded with recording from synaptically connected pairs of retina ganglion cells and geniculate neurons. In this situation, fractal exponents for retina and target neurons in the lateral geniculate body were nearly identical. This was interpreted to mean that fractal behavior is either transmitted across synapses, or has a common origin for the synaptically connected pre- and postsynaptic structure. On the other hand, fractal activity of medullary sympathetic premotor and the synaptically connected pregangionic synmpathetic neurons is apparently generated independently (Orer et al, 2003).

More support for the notion that ion channel properties play an important role for determining neuron performance comes from demonstrating a kind of memory mechanism for traces of prior activity in voltage-gated Na channels (Toib, 1998): time constants of channel recovery stand in a power function relation to duration of prior activation. The question of primary interest is of course how the dynamics of ion channels relates to the functional characteristics of a whole neuron. Gilboa et al (2005) addressed this question in a computational model of an ensemble of ion channels. In analogy to a 'real' neuron, this model neuron exhibits various dynamics at different time scales: a power law function recovery time scale after stimulation, temporal modulation of discharge pattern during maintained stimulation, and the dependence of adaptation to a stimulus step on the duration of the priming stimulus. The suggestive implication is that the ensemble of ion channels can exhibit properties on many scales, comparable to 'real' neurons, thus supporting the notion that the 'macroscopic behavior' of the 'real' neuron is, in fact, the result of cooperative fractal channel kinetics.

In addition to the studies cited in foregoing paragraphs, there are numerous reports documenting fractal-rate behavior in single neuronal point processes. However, these data were generally obtained for examining spike trains for encoding stimulus properties, and they are quite heterogeneous as regards species, neural structure examined, use of anesthetics and experimental conditions. Although this imposes serious limitations on drawing inferences on general principles, I select here a few studies which applied several of the commonly agreed upon and typical indicators of fractal properties, such as self similarity of firing rate with different averaging time, increase of spike number variance-to-mean ratio with counting time, and power law scaling relating the variable of interest to the resolution of measurement. In a series of publications, Grueneis et al (1993) reported fractal properties in spike trains recorded under various conditions including REM sleep of cats. In visual cortical areas of cats and macaques, Baddeley et al (1997) observed consistently non-Poisson spike train statistics, with some displaying self-similarity. Other neural structures examined included medullary sympathetic neurons (Lewis et al, 1993) and dorsal horn of the spinal cord (Salvador and Biella, 1994). A common feature of these and other like reports not cited here, was the lack of agreement on a consistent mathematical model that would satisfactorily describe the fractal process underlying the experimental data. In a study of retina ganglion cells, Teich & Saleh (1981) suggest a shot-noise driven self exciting point process; in a later study of the same experimental object, Teich found a modulated gamma-r-renewal process satisfactory while Grueneis et al(1993 ) favor a clustering Poisson process. Mandelbrot and van Ness (1968) considered Fractal Brownian motion as candidate. Clearly, the goal of determining whether a common principle governing spike train variability could be identified, and if not then for what reason, eluded this group of investigators.

Without examining specifically for manifestations of fractality, a number of investigators attempted statistical characterization of neural point processes, primarily motivated to reconcile irregularity of spike trains with their presumptive function as "code" of neural signals. But recall that Harris (2005) attributes irregularity of spike train discharges to cell assembly organization. In various modifications, the general approach chosen by Sakai et al (1999), Cateau and Reyes ( 2006), and Feng and Zhang (2001) consisted in designing model neurons to generate spike trains whose statistics would match that of "real" neurons recorded in animal experiments. Shinomoto et al (2003) recorded spike sequences from different cortical areas in awake macaques which they classified phenomenologically into different groups. Salinas & Sejnowski (2002) and Stevens and Zador (1998) assigned the principal source of discharge variability to correlations in the input feeding the examined neuron. None of these results warranted the allocation of observed or simulated spike train data to one of the probability distributions in the conventional repertoire of statistics, but Maimon and Assad (2009) at least excluded Poisson – like randomness from being a universal feature of spike time distributions in primate parietal cortex. In an exquisitely elegant experiment, Evarts (1967) followed the changes of interspike interval (ISI) histograms in premotor cortex pyramidal neurons in wakefulness, sleep and the phase of sleep associated with low-voltage fast EEG. Regrettably, his characterization of the ISI histograms is limited to rejecting Poisson distributions. However, inspecting the histograms displayed in Fig. 12 of his publication arouses one's suspicion of a long-tail distribution for sleep activity.

In a notable and very extended comparison of cortical neuron discharges in alert macaques, Shadlen & Newsome (1998) attributed to single neurons the ability to perform simple algebraic operations resembling averaging by combining inputs from several sources but they cautiously concluded that irregularity of the interspike interval distribution precludes them from reflecting information about the actual temporal structure of the synaptic input. They rejected random walk models of the kind applied by Gerstein and Mandelbrot as inadequate for capturing the statistical features of spike interval distributions, and found Poisson and various renewal processes likewise failing to yield satisfactory and consistent correspondence with recorded data.

If there is one conclusion that can be drawn from the extant data on the statistics of spike interval distributions, then it is that demonstrating fractal properties in spike trains requires carefully selected conditions. Multiple convergences from incoming pathways appears to obscure characteristic statistical properties of discharges in the recipient neurons. Thus, a neuron's intrisic connection pattern carries the burden of discharge variability. This view is reminiscent of Harris's (2005) view that spike discharge variability may be a signature of cell assembly organization. This is perhaps also the source of futility of assigning any information bearing capacity to discharge patterns of individual neurons (see for instance: Werner, 2007a). On the other hand, the more direct a neuron's connection pattern to peripheral sensors is, the more distinctly are fractal discharge properties demonstrable. But the opposite also seems to be the case when neurons are embedded in a network, as the observations of El Boustani et al (2009) in Section 2.3 show. The place to look for consistent fractal properties is apparently the macroscopic, global level of brain organization (see Section 2.3). Whether and how the mesoscopic level of the next Section bridges bridges the gap is the subject of the next section

## 2.3 The mesoscopic level of organization

Despite their relative simplicity, in vitro cultured neuronal networks are here viewed as mesocopic in the sense of representing neuron ensembles which exhibit rich spontaneous dynamical activity in the form of periodic bursting (Robinson et al, 1993; Nakanishi and Kukita, 1998; van Pelt et al, 2004; earlier references are cited in: Beggs and Plenz, 2003, 2004 ). At superficial inspection, the brief burst of activity lasting tens of milliseconds are separated by quiescence lasting up to several seconds (Corner et al, 2002; Tateno et al, 2002). The spontaneous emergence of patterns in the discharge trains was also noted by Giugliano et al (2004) and replicated in computational models . In extension of earlier work that led to identification of scale invariant Levy distributions and long range correlations in cultured neuronal networks (Segev et al, 2002; see Section 4 ), Segev et al (2004) attributed the activity bursts to be associated with the formation of statistically distinguishable subgroups of neurons , each with its own distinct pattern of interneuronal spatiotemporal correlations. Wagenaar et al (2006 a) emphasized the extremely rich repertoire of bursting patterns during the development of cortical cultures. The cultured cortical networks spontaneously generated a hierarchical structure of periodic activity with a stereotyped population-wide spatiotemporal structure. These recurring patterns ( called by the authors 'superburst' ) converged periodically to a dynamic attractor orbit, and were taken to imply large-scale self-organization of neurons *in vitro* , refuting the commonly held view of having random connectivity (Wagenaars, 2006 b). The recorded data of these authors are available for distribution to interested investigators. Departing from the practice of focusing on spontaneous activity, Breskin et al (2006) explored the propagation of stimulus evoked activity in neuron cultures. A graph theoretic analysis of their data attributed the dynamic evolution of the network connectivity to a percolation transition with power law characteristics, while the degree distribution of the grown network failed to meet power law criteria.

Working with mature organotypical cultures and acute slices of rat cortex, Beggs and Plenz (2003, 2004) concluded that the cascades of propagating neuron discharges they observed were indicative of the neural culture being in a state of self-organized criticality. The importance of this claim, and a recent disputes of its validity (see below), warrant close attention to methodological details: Beggs and Plenz based their analysis on recording spontaneously appearing negative local field potentials (NLFP), apparently occurring preferentially in cortical layers 2/3 (Gireesh and Plenz, 2008). The peaks of NLFP were considered indicative of synchronized population discharges occurring near the recording electrode tip (Plenz and Aertsen, 1993), on the rational that bursts of multiunit activity are more likely to generate large NLFP's than are single neuron discharges (Plenz and Thiagarajan, 2007). Accordingly, brief bursts of synchronized action potentials were the units of analysis in their experiments.

The alternation between brief burst of NLFP activity and quiescent periods remained in their experiments stable with a high degree of temporal precision, extending over periods of many hours. Beggs and Plenz set out to examine the idea that these cascades of neural activity may constitute a special mode of network activity, possibly of the character of "avalanches", indicative of SOC (Bak et al., 1987). They had specifically in mind the kind of self-organizing branching process discussed by Zapperi et al, ( 1995) and de Carvalho and Prado (2000). To examine this idea required determining the statistical properties of the observed activity patterns. For carrying out this analysis, they first defined the spatial pattern of signal-carrying electrodes during one time bin as frame; a sequence of consecutively active frames, preceded and ended by a blank period was called an avalanche. The NLFP did not propagate in the network in a spatially contiguous manner, thus excluding wave-like propagation. With these definitions and precautions in place, distributions of avalanche size and lifetimes were found to scale in cultures and acute cortex slices with a power law exponent -3/2, this value being resilient to various choices of scales (Plenz and Thiagarajan, 2007). The branching parameter was determined as sigma =1.04 . Being statistically indistinguishable from the ideal value of 1 , it signifies a critical state in the sense that activity starting at one electrode would initiate activity in one other electrode, on the average, keeping the network at the edge of criticality (Harris, 1989). This complements the evidence for fractal properties and supports the validity of the working hypothesis Beggs and Plenz started out with: to view avalanches as manifestations of the collective critical dynamics of SOC. More recently , fractal scaling of patterned neural activity was reported to also occur in cultivated neurons of leech ganglia and rat hippocampus (Mazzoni et al, 2007) ; and Pasquale et al, (2008) describe avalanches in dissociated neuronal cultures of cortex from embryonic rats.

Avalanches were subsequently also studied in superficial layers of rat prefrontal cortex (Stewart and Plenz, 2006) and during development of cortical layer 2/3 where they display nested theta- and beta/gamma oscillations ( Gireesh and Plenz, 2008). The theory of critical states predicts (see Section 2.4.3) and experiments confirm that neuronal avalanches display a maximized dynamic range of responses to stimuli (Shew et al, 2009). Functional architecture of avalanches conformed to Small World Topology (Pajevic and Plenz, 2009). Comparing NLFP activity *in vitro* cortex preparations with *in vivo* activity of awake macaque monkey cortex, Petermann et al (2009) established that high fidelity propagation of local synchronized scale-invariant activity patterns is a robust and universal feature of cortex in awake monkeys. Furthermore, large amplitude negative field potentials (like those constituting the avalanches) spread in a cascade-like fashion through the cortical network without distortion, much like action potentials: Thiagarajan et al (2010) described these stereotypical waveforms as 'coherence potentials'. They occur often in rapid succession as a stream of dynamical associations, suggesting the switching of the cortical network from one dynamical state to another.

The reason for viewing neuronal avalanches as manifestation of self-organized criticality was based on their fractal scaling properties for size and duration. Here just a brief reminder of the amply documented 'family resemblance' of fractality and SOC, to which the publications of Grinstein (1995), Chen et al (1995), Paczuki et al. ( 1996), Tebbens and Burroughs (2003) and Cessac (2004) speak, as do the model computations of Papa and da Silva (1997) , da Silva et al (1998), De Arcangelis et al (2006) and Levina et al (2007. However, despite the 'avalanche' of research on mechanisms leading to scale invariance, there exist questions about the necessary conditions for establishing the self-organized critical state (Kinouchi and Prado, 1999). A non-conserving critical branching model was proposed by Juanico et al,(2007) to demonstrate that SOC can occur in mean-field sand piles, provided the branching process is coupled to a background activity of spontaneous switching between refractoriness and quiescence among system components; in the stationary state, the system can undergo a transition from a subcritical to a critical state. In an elaborate recent study, Bonachela and Munoz ( 2009) claim that non-conserving (dissipative) dynamics does not lead to *bona fide* criticality. Non-conserving systems are in their view not truly critical models. Instead, non-conserving systems (such as the brain) would just hover around a critical point (presumably after some form of fine-tuning) with broadly distributed fluctuations which do not disappear at the thermodynamic limit. Such systems could be fluctuating in the vicinity of the critical point, but not at it. The authors call this condition 'self-organized quasi-criticality'. Whether this is the last word in the long standing debate of conditions for criticality in SOC remains yet to be seen.

Conditions for universality of 1/f scaling in dissipative self-organized criticality models were established by De Los Rios and Zhang (1999). Models predicting avalanches of neural activity include the work of Herz and Hopfield (1995) and were noted by Eurich et al (2002) in a network of globally coupled nonlinear threshold elements. Models of neural networks of non-leaky integrate-and-fire neurons exhibit over a wide range of connectivity patterns power law avalanches with an exponent closely approximating that reported by Beggs and Plenz for tissue cultures (Levina et al, 2007). In a comment to this paper, Beggs (2007) gives a lucid account of how neural networks could self-organize to operate at criticality. Critical avalanche networks can be computationally constructed by simple network growth models (Abbott and Rohrkemper, 2007). An exponent of the experimentally determined value -1.5 of avalanche size and lifetime scaling is predicted by the neural field theory of Buice and Cowan (2007).

A field theoretic approach was also investigated by Freeman (2005): analysis of high resolution electroencephalograms of rabbits revealed neural fields in the form of spatial patterns in amplitude and phase modulation of gamma and beta carrier waves which distinguished positive and negative conditioning stimuli. The goal of applying field theory was in these experiments to model states and state transitions as large-scale spatial patterns of neural activity for quick changes in adjustment to different behavioral situations. The cortical states were viewed as "wave packets", resembling frames in motion picture, stabilized in a scale free state of self-organized criticality. Recall, however, the reservations raised by Touboul and Destexhe (2009) that substantiating interdependence of fractality and self-organization requires additional evidence. In an entirely different context (namely fossil extinction), Newman (1996) shows that fractality need not be a unique indicator of SOC and criticality since an alternative simple model can account for the empirical power law relation (for an extended discussion of this, see Sections 4 and 5)

Seeking to strengthen the conjecture of self-organized criticality of avalanches, Plenz and Chialvo (2010) acquired experimental evidence that the neural avalanches in superficial layers of cortex exhibit five additional quantitative aspects of their dynamics which are consistent with critical dynamics. These were: separation of time scales between triggering and the avalanching event itself; stationarity of size distribution; temporal clustering prior to and following large avalanches , resembling Omori's law; power law decay for avalanche size in the wake of preceding large avalanches; and a fractal dimension for scaling spatial spread. The importance of these results lies first, in affirming evidence for avalanches displaying robust critical behavior; and second, in suggesting that their scale-invariant (fractal) properties do in fact reflect cortical networks being in a state of criticality. This is also supported by the observation that the exponents of simulated branching processes at near-critical branching are similar to scaling exponents characterizing oscillations in the MEG recorded alpha frequency band of humans at rest (Poli et al, 2008).

Criticality in cortex will be take up in Section 2.4.3, but let me merely stress at this point that it implies the brain being poised to undergo sudden second order phase transitions to new configurations with long range correlations among its disparate constituting elements. The dynamics is universal in its independence of details at the microscopic level (Sornette, 2000; Stanley, 1987). Viewing neural activity in this framework is a fundamental departure from a wave-type oscillatory or stochastic dynamics as the more commonly considered theoretical approaches in Neuroscience. To underscore this distinction, early-stage activity in the developing retinal network can be cited as an illustrative example (Hennig et al, 2009) of the essential feature of critical dynamics: rhythmic bursts of action potential in retina ganglion cells , propagating as wave-like events across the retina surface, arise at a very specific network state which meets the criteria of the classical percolation model of statistical mechanics (Essam, 1980): the phase transition consists in separating states of purely local from global functional connectedness, the latter displays in addition conspicuous fractal properties (Stauffer and Aharony ,1991/1994).

It may be revealing to contrast the failure of consistently finding fractal activity patterns in individually sampled neurons (other than those receiving relatively direct input, see Section 2.2.2) with the abundance of fractal patterns of (mesoscopic) neuron ensembles. It raises the question whether the origin of the latter may be a matter of assembly organization: note that in the records of neuron cultures, it is the concurrent activity of interconnected neurons sampled by different electrodes that forms the fractal pattern; this is of course quite different from the sampling of neurons in point process analysis, guided by chance encounters of a microelectrode with one active neuron at the time. The puzzle posed at the end of section 2.2.2 thus finds perhaps its resolution in network architecture: Teramae and Fukai (2007) describe a model that shows how the fractal property of a few individual neurons can turn into an organized communal property of an ensemble. This lesson can also be learned from models of dynamic pattern formation in neuron populations, forming fractal power spectra and power law pulse distribution (Usher and Stemmler, 1995). Similarly, network amplification of local fluctuations causes fractal firing patterns and oscillatory field potentials in two-dimensional models of leaky Integrate-and-Fire neurons ; feedback connectivity of local excitation and surround inhibition being the essential prerequisites ( Usher et al., 1994).

Bedard et al. (2006 b) accept existing claims for 1/f scaling of global variables of neural activity (e.g. EEG: Freeman et al, (2003); EMG: Novikov et al., (1997); see Section 2.3) , and acknowledge them as evidence for self-organized critical states with power law distributions , much as the models of De Arcangelis et al (2006) and Levina et al (2007) suggest. They also accept the validity of claims for fractality discussed in Section 2.2.2 for various point process analyses. But they contest the legitimacy of generalizing from these disparate sources of data. The connection between 1/f frequency scaling of global variables and critical states of neural activity is, in their mind, far from firmly established. Moreover, they emphasize that the association of 1/f spectra with criticality may not be obligatory (for a review: see Giesinger, 2001; Newman, 1999).

Having raised this warning flags, Bedard et al (2006) not only confirmed in their own investigations the presence of 1/f frequency scaling in EEG of cat parietal cortex (in absence of anesthesia), but showed in addition 1/f frequency scaling in bipolar records of Local Field Potentials (LFP). That bipolar LFP recordings sample relatively localized populations of neurons was shown by Destexhe et al. (1999). They turned then their attention to investigating whether this 1/f scaling reflects self-organized critical states with the result that avalanche size distributions did in their experiments not follow power law scaling, nor did interspike interval distributions of concurrently recorded single neuron activity show 1/f scaling ; rather, the distributions were consistent with a Poisson process. Accordingly, Bedard et al. reject the evidence for critical state dynamics. Instead, they proposed a model that would account for 1/f frequency scaling without being associated with critical states. Their model shows that the observed 1/f scaling can indeed be produced by a band pass filtering effect of extracellular media. This means that extracellular field potentials (such as LFP) can show power law scaling while the underlying neuronal activity *per se* need not be critical. El Boustani et al (2007) found in experimental data and models of cortical “activated” states also evidence for Poisson spike distributions, and absence of avalanche dynamics. The irregular states of cortical networks are thought to stem from a very high dimensional deterministic chaos. However, as alternative, it is worth recalling that Harris (2005) views spike train irregularity as one of the signatures of cell assembly organization.

Investigating further the discrepancy between the Beggs-Plenz and the Bedard et al observations, Touboul and Destexhe (2009) recorded avalanches in cortex of (awake) cats, paying careful attention to the conditions of data collection: avalanche statistics of negative peaks LFP (linked to neuronal firing), positive peaks (unrelated to neuron firing) and surrogate data (obtained by random shuffling experimental data) were analyzed; avalanche criteria were those of the Beggs-Plenz studies. Essentially, time and amplitude distributions of NLFP showed power law distributions, preferably at high detection thresholds. But, positive NLFP and surrogate data (randomly shuffled peak times -- essentially equivalent to a threshold stochastic process-- can also show power-law distributed amplitude distributions. The conclusion of this study, then, was that apparent power law scaling cannot be considered as proof of self-organized criticality .

At this point, a comment seems in order: the publication of Gireesh and Plenz (2007) seems to suggest that cortical layers 2/3 are the preferential site of avalanche occurrence. The data analyzed in the Touboul and Destexhe study were obtained in earlier work of Destexhe et al (1999) in parietal cortex of awake cats; there is no indication that a possible role of cortical layer was considered. Whether layer specific patterns of neuronal arborization (Callaway, 2002) could be a source of the discrepancy can at this point not be determined

The discrepancy between presence of 1/f scaling in ensemble neural activity (EEG and LFP) and, yet, 1/f scaling not being an intrinsic property of the individual neuron itself, that Bedard and El Boustani et al claimed to have identified, motivated El Boustani et al (2009) to adopt yet another experimental approach: to this end, they measured the scaling properties of the power spectrum of the intracellularly recorded membrane potential of individual neurons. The experiments were conducted in cat primary visual cortex *in vivo* , the animals being anesthetized and paralyzed. Full-field visual stimuli of varying characteristics were presented to the dominant eye to drive the cortical region under study to states with different firing characteristics. The remarkable result was that the frequency scaling of individual cells was largely determined by the visual stimulus statistics. There was no consistent relation between individual neurons' scaling exponent and the visual stimuli, neither was there any correlation between membrane potential and the spiking scaling exponents. Various control tests and a computer model corroborate the authors' conclusion that statistical correlations in a neuron's input (i.e. its presynaptic activity) can modify the power law exponent of its spiking activity. Hence, it appears that modulation in a neuron's power law exponent may reflect changes in the correlation state of the network activity. According to these findings, intrinsic cellular properties do not seem to play a major role for its scaling which reflects in the authors' s view primarily the network context.

Regarding self-organization, El Boustani and Destexhe (2009) follow the lead taken in Destexhe's Doctoral Thesis of 1992 and observations of Korn and Faure (2003), and present new evidence in support of chaotic dynamics in EEG: sensitivity to initial conditions is of course prominent; it is also associated with broad-band power spectra and a fractal attractor dimension. The authors confront at length the puzzle that coherence and low dimensionality at the macroscopic level of EEG is associated with stochastic neuronal dynamics at the microscopic level. Is this comparable to conditions obtaining in thermodynamics ?

Whence criticality ? In peripheral neurons, it seems to be favored by closeness to input from peripheral receptors (Section 2.2.2). At the mesoscopic level, Plenz and Chialvo's ( 2009) analysis of avalanches in primate cortex seem to assure legitimate criticality at the mesoscopic level; yet, the work of Bedard et al (2006) and El Boustani et al (2009) raises the possibility that scaling properties of neuron activity may not be of intrinsic neuronal origin, but a consequence of network activity. The next Section will continue to ask: if and where in the nervous system, and under what conditions, does fractality and criticality in the brain originate ?

## 2.4 The macroscopic level of neural organization:

### 2.4.1: Fractals in brain networks

Fractality at the macroscopic brain level should be viewed in the context of, and in reference to, the two major conceptual and observational frameworks that have come to guide neuroscience research: the network structure of cortical connectivity, and the brain's state of criticality resulting from the complexity of nonlinear dynamic interactions among its constituents. Advances in network theory (Albert and Barabasi , 2002; Dorogovtsev, 2002; Park and Newmann, 2004 ) influenced the application of computational and graph-theoretical methods for characterizing structural brain connectivity in accord with statistical and topological criteria (Hilgetag et al 2002). Examining the columnar organization of neocortical cortex in detail, Roerig and Chen (2002) found that the number of connections to a central neuron has the shape of a long-tailed histogram, fitting a power law. On the basis of this "bio-power-law connection probability function", Stoop and Wagner (2007) tested a range of network types for spread of synchronization among cortical columns: the superiority of the power-law connection was evident. In general, interaction among neurons and neuron ensembles by synchronization is constrained by network topology (Arenas et al, 2008), hence the relevance of network architecture for Neurodynamics. The potential role of neural synchrony for perceptual organization and conscious experience is a subject of a recent review by Uhlhaas et al (2009). There is considerable evidence that anatomical and functional connections between different cortical areas possess an intricate organization in the form of "small world networks" (Watts & Strogatz, 1998) , forming clusters of nearby cortical areas with short links, which in turn have long range connections to other clusters (Hilgetag and Kaiser, 2004; Sporns and Zwi, 2004; Sporns et al, 2004; Stam, 2004; Stam and Reijneveld, 2007). Neuroanatomical data sets permit identifying a repertoire of characteristic structural building blocks (motifs) (Sporns and Koetter, 2004).

A hierarchical cluster architecture is thought to provide the structural basis for stability and diversity of functional patterns in cortical networks (Kaiser et al., 2007; Kaiser, 2008).

Moreover, hierarchical modular topologies assure sustained activation in neural networks, intermediate between rapid fading and generalized activity spread. This can be considered a prerequisite for the occurrence of criticality (Kaiser and Hilgetag, 2010). Hierarchical graphs can switch between different dynamic activity patterns, depending on the level of ongoing (spontaneous) background activity (Muller-Linow et al, 2008; Hutt and Lesne, 2009). In terms of hierarchy theory, these investigations do not specifically address the implications of nested hierarchies which, however, are suggested by the finding of inter- and intra-cluster network hubs (Sporns et al, 2007). Moreover, fMRI data obtained from subjects in resting state identify strong functional connections between regions for which no direct structural connections are known (Honey et al., 2009). This finding may be an indicator of nested clustering (see Sections 2.4.3 and 6).

In the absence of deliberate external stimulation, neuronal cortical dynamics displays complex spatial and temporal patterns of activity. In simulations of networks that mimick the large-scale inter-areal connection patterns of cortex, activity takes place spontaneously at multiple time scales, punctuated by episodes of inter-regional phase locking of oscillations (Honey et al, 2007). Significantly, the connections link neural populations of multiple levels of scale, from whole brain regions to local cell columns: this suggests that cortical connections may be arranged in fractal, possibly self-similar patterns. Statistical measures of a computational model of a fractal connection pattern did in fact resemble those of a real neuroanatomical data set (Sporns 2006). The computational models also show that varying fractal patterns induce strongly correlated changes in several structural and functional measures of network properties, as evidence of their interdependence.

In general, scale free complex networks display self-similarity under length-scale transformations (Song et al, 2005) but not necessarily with regard to degree distribution (Kim et al, 2007), but models of scale-free networks need not necessarily be fractal. How, then, can the fractality of many naturally occurring networks come into being ? Song et al. (2006) account for the simultaneous emergence of fractality, modularity and small-world effect, as well as the scale-free property of real world networks by a multiplicative growth process: the network growth dynamics is conceived as the inverse of a renormalization procedure, whereby the network hubs accrete connections by linking with less connected nodes, which leads to a robust fractal topology.

Within the small-world network clusters, functional Magnetic Imaging (fMRI) identifies a scale-free connection pattern inasmuch as the number of links per network node (the node degree) satisfies a power law relationship (Eguiluz et al, 2005). Likewise, van den Heuvel et al (2008) find In an imaging study of the resting brain, that inter-voxel connections follow power law scaling as evidence for scale free network topology, possibly associated with a small-world organization. This form of organization is associated with conserved wiring length and conducive to synchronization of activity across the network (Zhou et al,2007; see also Changizi, 2003; Section 2.1).

Although citing merely a small fraction of the numerous publications concerned with relations between network topology and dynamics, this section underscores two points of relevance for the objective of this review: first, the presence of, and effect on network dynamics of hierarchic network organization (itself being of several types); and, second, effects of network fractality on network dynamics; but the functional implications of the latter, notably for criticality, require further investigation, as does the possibility of self–similar modularity in brain networks. In the case of metabolic networks, the latter is shown to affect path connections for diffusion and resistance of flows (Gallos et al., 2007).

### 2.4.2: Fractals and Criticality of Brain States

Criticality, listed in the foregoing as the second notable feature in current thinking about global brain function designates the view that brain is under normal circumstances at the verge of undergoing a second order phase transition. This is attributed to its complex organization of a large number of components interacting via nonlinear dynamic functions.

Measuring the fractal dimension of EEG records, Babloyantz (1986) related different values with differences in sleep states. With subjects acting as their own controls, inhalation anesthesia causes a noticeable increase in EEG dimensionality (Mayer-Kress and Payne, 1987). Multichannel MEG records, obtained with a SQUID show scaling with varying degrees of scale similarity , decreasing with the distance between recording channel locations (Novikov et al, 1997). Studying dynamical synchronization in the brain, Gong et al (2003) find scale invariant fluctuations of dynamical synchronization in human EEG. Linkenkaer-Hansen et al (2001) report long-range temporal correlations and scaling with 10-20 Hz brain oscillations. Pursuing this observation in more detail, Linkenkaer-Hansen et al (2003, 2004) suggest that the long-term spatial-temporal structure of the complex ongoing EEG activity may reflect a memory of the system's dynamics extending beyond just a few seconds, possibly by a continuous modification of functional brain networks in the sense of SOC . In these tests, somatosensory stimuli attenuate temporal correlations and power law scaling behavior, suggesting that stimuli degrade the network memory of its past. The relationship to SOC was also the subject of the work of Freeman et al (2003) in measurements of temporal and spatial power spectral densities that identify EEG phenomena as fractal. Moreover, Freeman (2005) proposed a field-theoretic approach to account for scale-free neocortical dynamics. In five frequency ranges (extending from 0.5 to 48 Hz), detrended fluctuation analysis of EEG show global synchronization time series with scale free features (Stam and de Bruin, 2004); the scaling exponent differs for conditions of eye open and eye closed. Stam (2005) also reviewed the nonlinear dynamical analysis of EEG and EMG at great length. Positive and negative feedback affect the scaling exponent of EEG differentially; this was determined in a detrended fluctuation analysis (Biuatti et al., 2007). Performance in Stimulus detection of weak stimuli is best accounted for by modulation of the power law component in the power spectrum of MEG record : Shimono et al, (2007) attribute this phenomenon to the brain operating in a state of self-organized criticality which modulates the power spectral exponent to optimize responsiveness to external stimuli.

Transients in EEG records can be detected as differences in fractal dimension of EEG (Arle and Simon, 1990), as can be neuropathological conditions (Paramanathan and Uthayakumar, 2008 ) , and differences in age and gender (Nikulin and Brismar, 2005). Nonlinear spectral analysis enabled Kulish et al (2006) to determine in EEG a set of generalized fractal dimensions and fractal spectra which reveal differences in subjects when replying to questions with either YES or NO. In a study of human development from infancy to 16 years of age, Thatcher et al (2009) measured phase shift duration and phase locking intervals of the EEG for computing instantaneous phase differences between pairs of electrodes; the log-log spectral plots showed 1/f distributions. The data revealed increased phase stability in local systems, paralleled by lengthened periods of unstable phase relation between distant connections. These results were taken to reflect progression towards self-organized criticality, accompanying the growth spurts from infancy to adolescence. When Listening to music Bhattacharya and Petsche (2001) find homogeneous scaling in the gamma band EEG over distributed brain areas, whereas the homogeneity is reduce at rest, or when reading text or during spatial imagination. As is well known, music has been under scrutiny for fractal properties for quite some time, see for instance : Voss, 1975; Hsu and Hsu, 1991; Boon and Decroly, 1995; see also below: Bianco,2007).

Long-range temporal correlations in spontaneous discharge patterns of hippocampal-amygdala complex neurons show a power-law relation in epileptic patients (Bhattacharya et al, 2005): activity of individual neurons was in this study recorded by means of micro-wire electrodes that had been implanted for localization of epileptic foci, and records were taken in inter-ictal periods, with the subjects being awake . Neuronal activity in substantia nigra exhibits fractal activity in anaesthetized rats, but was strikingly absent in the dopaminergic nigrostriatal neurons with relatively constant discharge rate (Rodiguez et al, 2003). The authors consider the possibility that pathological rhythmic discharges and tremor onset may be associated with loss of the fractal pattern of nigrostriatal neurons. During paradoxical sleep and in the attentive state, neurons in the mesencephalic reticular formation of unanaesthetized cats exhibit firing patterns with 1/f spectral profile (Yamamoto et al, 1986). Kodama et al, 1988) extended this observation to discharge properties of neurons in Hippocamus and ventrobasal thalamic neurons, and suggest that 1/f structured patterns in discharge trains are indicative of spatial and temporal summation of convergence. Variations in 1/f spectra in cortical and subcortical brain structures of monkeys are apparently related to differences in emotional states (Andersen et al., 2006).

In a very detailed and thorough study, Bianco et al (2007) identify the EEG time series as a (non-ergodic) renewal non-Poisson process, reflecting strong deviation from exponential decay. This startling claim is based on two premises: one, the comparison with the statistics of an entirely different physical process, namely the fluorescence intermittency in blinking quantum dots (Bianco et al, 2005); and, second, on the conjecture of the brain operating at or near a self-organized critical state. The implication is that neuron synchronization can be viewed as a kind of phase transition involving the close cooperation among many constituents of a neuron set, each individual neuron in essence losing its identity. Furthermore, the absence of exponential truncation would violate the ergodic condition (Bel and Barkai, 2005). The authors then proceed to show that compositional music belongs to the same category of processes. They finally claim that the effect of music on the human brain is in fact based on the essential identity of their respective fractal dynamics, ensuing a kind of complexity matching of the interacting brain-music systems. This aspect will be further pursued in section 5.1. Equally consequential are the inferences drawn by Allegrini et al (2008) from their EEG data. The thrust of their analysis is on measuring the time distribution of recorded events occurring simultaneously at two or more electrodes (in their terminology: coincidences); they find that the time interval between two consecutive coincidences has a waiting time distribution corresponding to perfect 1/f noise. The theoretical analysis of this finding leads these authors to infer that the coincidences are driven by a renewal process.

The electroencephalographic findings in support of 1/f scaling are supplemented by observations with brain imaging: In 1997, Zarahn et al (1997) reported BOLD time series data obtained from normal subjects at rest that exhibited a fractal power spectrum and self-similar signal contributions, with disproportionate contribution of power in the spectrum for low frequencies. The temporal variablility of brain activity in time series of fMRI data in combination with a voxel-wise analysis of scaling exponents enabled Thurner et al (2004) to distinguish different physiological states of the brain. In non-active brain regions, the voxel-profile activity is described by a random walk model; in contrast, stimulus activated brain activity is characterized as correlated fractional Browninan noise. The same group of investigators (Shimizu et al, 2004) examined fMRI time series with a multifractal method to extract local singularity (fractal) exponents: the range of Hoelder exponents in voxels with brain activation is close to 1, whereas exponents in white matter and voxels in the absence of brain activation are close to 0.5

Without further discussing at this point the far reaching implications of the non-ergodicity claim (Tsallis, 2009; Tsallis, et al, 1995), I merely alert to two publications which interpret human EEG signals in terms of a Tsallis Entropy measure (Capurro et al 1998, 1999).

The common theme of studies surveyed in the following is wavelet based representations of functional magnetic imaging (fMRI) time series. Amongst others, Wornell (1993) explicated in detail the role of wavelet based representations for the power law family of processes. The remarkable feature of wavelet analysis is that it can be viewed as matching self-similar processes since the wavelet coefficients exactly reproduce, from scale to scale, the self-replicating statistical structure of such processes (Abry, 2003).

Publishing with various associates since 1994, Bullmore gathered extensive experience with fractal analysis of human brain activity which led eventually to the suggesting that wavelet-based f MRI time series estimates (Bullmore et al, 2001) can be viewed as realizations of Fractional Brownian Motion , i.e. a class of fractals described by Mandelbrot & Ness (1986), characterized by zero-mean, and non-stationary and non-differentiable time functions (see Section 2). Extolling further the virtues of wavelet techniques for the purposes on hand, Bullmore et al (2004) and Maxim et al, 2004) give a meticulous account of their use of the 'discrete wavelet transform' approach to fMRI time series evaluation. In normal subjects at rest, the time series is most parsimoniously described as Fractional Gaussian Noise, signifying a persistent long-memory fractal processes of which the Hurst Exponent is a defining parameter. Interestingly, the value of this parameter in Alzheimer subjects differs from the norm (Maxim et al, 2004). Several results from the same laboratory contribute additional facets to the notion of the active brain displaying fractal properties: Achard et al (2006, 2008) applied discrete wavelet transform analysis to fMRI time series to estimate the frequency dependence of functional connectivity between some ninety cortical and subcortical brain regions; the functional networks is dominated by a neocortical core of highly connected hubs with an exponentially truncated power law degree distribution. Dynamical analysis of brain at wavelet scales from 2-37 Hz show the emergence of long-range connections with execution of motor tasks (Bassett et al, 2006). Under certain conditions (e.g. age, cognitive performance, certain pharmacologic interventions) brain dynamics requires a more comprehensive description than is captured by the monofractal analysis applied in the studies cited thus far (Suckling et al., 2008. In such cases, a more comprehensive description must make allowance for scaling behavior that is governed by several local scaling exponents. Multifractal analysis can then be characterized by the histogram of the Holder. Following expenditure of cognitive effort, the brain's fractal oscillations require several minutes for returning to baseline activity, this time depending on the task's cognitive load; this is taken to signify the relevance of fractal scaling for adaptive task processes, in addition to the role it plays for the "resting" brain (Barnes et al, 2009). The substantial evidence for modular organization of brain networks is reviewed by Bullmore and Sporns (2009), and was subsequently further refined by Meunier et al (2009), applying a method for rapid, high-resolution modular decomposition of brain functional networks (Blodel et al, 2008). Differences between low frequency BOLD signal spectral power in task and rest periods also support the notion of fMRI reflecting meaningful brain states (Duff et al, 2008), as do the emotional task dependent fractal fluctuations in fMRI of the cerebellar vermis (Andersen et al, 2006). Brain imaging in Neuropathology has revealed significant differences between patients suffering from unawareness of ownership of one arm (Asomatognosia) and those with additional confabulations (Somatoparaphrenia): the latter patients display lesions in the medial and orbitofrontal regions, in addition to the multiple large lesions including temporo-parietal sectors which are common to both groups of patients (Feinberg et al, 2010). For the distinction of stimuli related to the self (i.e. self-referential stimuli) from those not so related, Northoff et al (2006) identify processes mediated by cortical midline structures.

Alternative approaches to the question of brain criticality have a distinguished history: sudden transitions between stable states of motor behavior are well known since the pioneering observations of Haken et al. 1985; Kelso, 1995). The transitions were interpreted as manifestations of metastability in the self-organizing nonlinear dynamic system of the brain, along the theoretical lines formulated in Synergetics (Haken, 1983). Criticality in brain and behavior was first mentioned by Kelso (1984) in a brief note. Using a superconducting quantum interference device (SQUID) sensor array, Kelso et al (1992) reported a few years later their observations of spontaneous transitions in neuromagnetic field patterns which occur at a critical value of a behavioral parameter: coherent states of both brain and behavior were captured by the spatiotemporal pattern of phase relations among participating components. This was considered evidence for the brain being a pattern forming system that can switch flexibly from one coherent state to another. Chialvo credites also Varela (2001) with the vision of brain large-scale dynamical properties. Locating cortical regions associated with such phase transitions of motor behavior, Meyer-Lindenberg et al (2002) showed that TMS can induce switches between two clearly defined and distinct motor behavior patterns. Additional new evidence of aspects of critical brain behavior accrued in rapid sequence in the following years in several forms: as the scale free connection pattern of cortical networks (Eguiluz et al, 2005; Chialvo, 2004; 2008) ; as result of the avalanche analysis of Beggs and Plenz (2003,2004) at the mesoscopic level; as coordination dynamics of large scale neural circuits subserving rhythmic sensorimotor behavior (Jantzen et al, 2008); and finally from two fMRI studies at the macroscopic level, which will be the subject of the following paragraphs.

Two virtually simultaneously published recent studies, using different experimental strategies, deliver seemingly firm evidence for brain criticality. Kitzbichler et al (2009) based their approach on the widely accepted view that many behavioral and cognitive states are related to coherent or phase-locked oscillations in transient neuronal assemblies (for a recent summary: Womelsdorf et al, 2007). The measures for determining phase synchronization between component processes were in their study the phase lock intervals (estimating the length of time a pair of bandpass filtered oscillations remain in phase synchronization), and the lability of global synchronization (informally analogous to the previously discussed avalanches). Applying these measures to functional MRI and MEG data recorded from normal volunteers at resting state demonstrated power law scaling of both pair wise and global synchronization. They then evaluated the performance of two models , both typically being used in nonlinear dynamics: the Ising and the Kuramoto (1984) model. Observed and model generated data were identical, provided the model system was in a critical state. Hence, the authors conclude that the brain must be in a critical state. Moreover, the critical brain dynamics obtained at frequency intervals ranging from 0.05-0.11 to 62.5-125 Hz, confirming criticality of the human brain network organization across its functional bandwidth. They consider therefore ‘Broadband Criticality’ as a characteristic property of the resting brain network functional organization

Although also using the Ising model as reference point for determining brain criticality, Fraiman et al (2009) followed an entirely different approach: the issue at stake in their study was to determine whether and to what extent the dynamics of the paradigmatic two dimensional Ising model at criticality displays features that correspond to patterns encountered in the imaging of (resting) brain networks. However, unlike most prior studies of brain dynamics cited in this Section, no prior assumptions on structural connectivity of brain regions were made. Instead, network connectivity was extracted from voxel correlations: thus, networks were here defined in terms of correlations among the activity at each location (voxel in the case of the brain, and lattice site in the Ising model). Prior investigations showed that the so called ‘resting state’ (absence of overt external stimulation) is subject to a Default Network Dynamics, reflecting balanced positive and negative correlations between activity in component brain regions (Fox and Raichle, 2007; Baliki et al, 2008); this is not the case under certain abnormal conditions (Baliki et al, 2008) . The result was that the dynamics of the Ising model at criticality, as captured by the correlation networks, exhibits average statistical properties which are identical to those observed in the brain networks at resting condition.

Among several other network characteristics that match critical Ising dynamics with brain dynamics was also the equality of the fraction of sites with positive and negative correlations, corroborating that the dynamics of the normally functioning brain at rest being near a critical point. In any case, the unequivocal answer to the question the investigators set out to answer was that networks derived from correlations of fMRI signals in human brains are indistinguishable from networks extracted from Ising models at critical temperature.

In an important next step, Expert et al (2010) investigated the large-scale dynamical properties of resting brain by examining more closely the character of the spatio-temporal correlations: considering three successive steps in spatial coarse graining, two-point correlation functions exhibit self-similarity; self-similarity in time was revealed by 1/f frequency behavior of the power spectrum. The condition of long range correlations in space and time presupposes a dynamical system at criticality; the strong correlations across large distances are indicators of highly integrated cortical states, with nearby clusters functioning in synchrony.

Apart from this principal conclusion of this study, the authors of this study also alert to a significant property of the brain networks which, as noted before, are extracted from the site-to-site temporal voxel correlations: obviously, equally oriented spins in the Ising model coalesce in large domains near the critical temperature where also nontrivial collective states emerge in the Ising model's otherwise regular lattice. Similarly, large regions of brain activate concurrently with deactivation of other regions. How does the brain self-organize to negotiate the dynamic balance between the extreme possibilities of total quiescence and explosive massive excitation? The authors refer to a discussion of this stability problem which was already noted by Abeles (1991). It motivates their question: is it necessary to confine brain activation to structural connections linking brain regions, as is customary in most current research ? (e.g.: Hagmann et al, 2008). Take the Ising model as example: there, a change of temperature can lead to the emergence of functional collectives, in the absence of preexisting structural connections. This leads Fraiman et al. to ask: might the brain, likewise, have this capacity, as basis of a kind of adaptive coordination dynamics of the kind envisioned by Kelso and Tognoli, (2007) and Tognoli and Kelso (2009) ?

Apparently, SOC, metastability and phase transitions constitute a nexus of intimately interrelated dynamic processes of which fractals and self-similarity are pivotal aspects.

### 2.4.3: Significance of Brain Criticality

In statistical Physics, systems operating at the critical point of transition between ordered and random behavior are metastable with respect to a set of control parameters, and are capable of rapid qualitative change in response to fluctuations of external input. For systems far from equilibrium most of the analytical and numerical methods of the 'classical' (equilibrium) theory appear to remain valid (Sornette, 2000 ). Moreover, dissipative (open) Hamiltonian System, such as the brain, have the capacity to form "strange" attractors whose boundaries and bases have fractal properties (Aguirre et al, 2009; see Section 4). At or near the point of phase transition, the systems exhibit complex patterns of fluctuations on all scales of space and time, as one the indicators of an impending phase transition; another is the slowing down of relaxation processes, associated forming long range correlations for efficient functional coupling among system components: both events are anticipatory signals of impending critical transitions (Scheffer et al, 2009). Fractal clusters formed by phase transitions can be characterized in terms of correlation length (Antoniou et al, 2000) which is associated with fractal scaling of clusters of correlated elements on all scales; as a result, any intrinsic scale before phase transition is de facto 'forgotten' (Stinchcombe, 1989). As a corollary, the system presents at the critical transition qualitatively new properties, requiring new macroscopic descriptors. The important feature of the organization following the phase transition is to form new objects with distinct properties. In physics, this is manifest as, for instance, the phase transition from ferro- to para-magnetism, or from water to ice (Stanley, 1999). Typically, one deals with a large collection of 'microscopic' constituents which, at phase transitions, arrange to a macrostate which displays qualitatively novel features and properties. The macrostate's new properties have no referent at the microscopic level, and require new descriptors: by way of illustration, think of hardness or liquidity in the ice-water example as descriptors of new physical properties, originating de novo upon phase transition. The properties described in the foregoing are universal in the sense that the apply irrespective of the system's constituents at the microscopic level.

One of the amazing features of phase transitions is that material systems of diverse physical properties at their microscopic level form on phase transition but a small number of Universality Classes which share identical macroscopic properties ( for a discussion in relation to brain function: see Werner,2009 c). Based on a stochastic theory of neural activity, Buice and Cowan (2007) developed field theoretic methods for nonequilibrium statistical processes; their model exhibits a dynamical phase transition of the universality class of directed percolation (see Section 2.4.2).

Critical Theory (Stanley, 1987; Marro and Dickman, 1999) considers reality as a hierarchy of levels, each having its own scale, its own description and a theory that accounts for that description. The scale on each level emerges from the scale on the next finer level by ignoring some of the lower level details which become invisible at the higher level scale (Laughlin, 2005; Sokal and Bricmont,2004)). The result is a drastic reduction of dimensionality. Coarse graining (specifically renormalization group transformation ) (Fischer, 1998) unveils self-similarity at the point of phase transition. The intimate relations between scaling, renormalization group, and long-range correlations are addressed by Perez-Mercader (2004) and Penrose (1986), the latter pointing out that the definition of fractal dimension depends primarily on the distribution of widely separated sites, telling little on sites that are close together.

What is the significance of criticality ? Excitable systems at criticality exhibit an optimal dynamical range for information processing (Kinouchi and Copelli, 2006; see Shew et al, 2009, in Section 2.3) . Furthermore, a model that reproduces the typical features of systems at a critical point learns and remembers complex logical rules: learning occurs by plastic adaptation of synaptic strengths, and exhibits universal features in being independent of the specific task assigned to the system (Arcangelis and Herrmann, 2010). Finally, Phase transition in critical systems provide an universal mechanism for rapid switching between different cooperative neuron collectives. These three attributes of criticality are the reason for its rapidly moving into center stage of current brain theory.

3. Psychological and Behavioral Processes

The following overview of psychological functions with power law scaling is predicated on the notion that mental states may be viewed as macrostates emerging from EEG dynamics (Allefeld, 2009), and neurophysiological processes generally. Classical Psychophysics of Helmholtz, Fechner and Weber sought to establish dependencies of perceptual experience on properties of physical stimuli impinging on sensory organs. In 1975, Stevens reported the summary of the extensive work that led him to propose that this dependency is in many sensory modalities a power function. In neurophysiological experiments, Werner and Mountcastle (1963,1964) identified the power function scaling of responses in primary afferent cutaneous nerve fibers to mechanical indentation of peripheral receptors. Neurons of primary visual cortex (V1) exhibit a higher coding efficiency and information transmission rate for input signals with natural long term (1/f) correlations (Yu et al, 2005). Copelli et al (2002) and Kinouchi and Copelli (2006) claim that Stevens' law (1957) for intensity of subjective sensory experiences can be attributed to dynamics in a network of excitable elements constituting the peripheral receptors, set at the edge of a phase transition, i.e.: of being in a state of criticality. For a discussion of this view, see Chialvo (2006).

Unlike dismissing the fluctuations in the performance of many psychophysical task as "noise", Gilden ( 1997,2001) attributes them to a memory process associated with active choice and discrimination. This memory process is suggested to express itself as 1/f 'noise' in the three major measurement paradigms in Psychophysics: speeded judgment, accuracy of discrimination and production. The 1/f fluctuations are attributed to an intrinsic dynamics, associated with the formation of representations, comparable to the kind of memory that arises in dynamical systems as they flow forward in time, along principles outlined by Beran (1994). According to this interpretations of the psychophysical observations, cognition would generate its dynamical signature as a consequence of its own activity: this would entail a fundamental revision of what is signal and what is noise in psychophysical data. Gilden points out that" the conventional experimental design and data analysis using ANOVA does in fact bury " one of the most important signatures of what happens when the mind is working".

Timing fluctuations in tasks requiring sensorimotor coordination display cycle-to-cycle fluctuations which, analyzed as time series, show fractal scaling of power spectra. Ding et al (2002) suggest that the reason for this lies in the multiple time scale activities of distributed neural areas that contribute to the task performance. If asked to produce random series of numbers from a given set, series with short and long range correlations are produced which in most cases exhibit a power law spectrum (Morariu et al, 2001). Van Orden et al (2003) interpret serial correlations in human cognition as evidence of self-organization. In their view, self-organization coordinates the activities of the organism across a hierarchy of time scales, producing correlated variation across time: variations in response times would then appear as a natural fractal in which larger scale deviations nest within themselves smaller (self-similar) scale deviations. Accordingly, 1/f noise is in this view not sufficient evidence for self-organized criticality, but rather its necessary consequence. Similarly, Kello et al (2007) assembled reaction time and response data which lead them to considering the 1/f scaling of their data as expression of a coordinative, metastable basis of cognitive functions. This view is in effect an extension of Van Orden's et al (2003), shifting the genesis of 1/f scaling from self-organization to metastability: the 1/f pervasiveness in the brain would be the signature of metastability associated with cognitive functions. However, these claims are challenged by Wagenmakers et al (2005) and contrasted with the alternative that long term serial dependence in data can be explained in a number of ways , for instance by mixtures of a small number of short-range processes) (Wagenmakers et al, 2004).

Applied to problem solving and insight, reasoning was viewed by Stephen and Dixon (2009) as the self-organization of novel structures: taking a particular problem solving task as example, the authors suggest that the problem solution can be viewed as a phase transition in a self-organizing system whose dynamics would be reflected in power law behavior. Implications for social psychology are reviewed by Correll (2008): cognitive effort to avoid bias in judgments reduces the scaling exponents of response times relative to less challenging tasks. Grigolini et al (2009) interpret Correll's data to suggest that increasing the difficulty of cognitive tasks would accelerate the transition from observed 1/f noise to white noise in decision making time series.

The temporal structure of many human-initiated activities can display a striking regularity. Barabasi (2005) showed that a decision-based queuing process can account for the dynamics of some human patterns of activity: when individuals execute tasks based on some perceived priority, the timing of the tasks will indicate the signatures of fractal dynamics: heavy-tailed distributions with initial fast bursts.

If patterns of expression in spoken language reflect in some way the organization of brain processes, then Zipf's law is of course the notable landmark that presages more recent fascinating reports of fractal patterns and scale-invariant word transition probabilities in spoken and written texts (Costa and Sigman, 2009; Altmann et al, 2009; Alvarex-Lacalle et al, 2007), and their extension to music (Zanette, 2008). On the basis of EMG data, it appears that some common features of patterning in language, music and syntax (Patel, 2003) can be attributed to neural activity in Broca's area and its right hemisphere homologue (Maess et al, 2001).

### 3.1 Symbol processing and fractals

The classical book "*Language of Thought*" (Fodor, 1975 ) epitomizes the framework of computation-representation of the Computational Theory of Mind. However, with adopting a dynamical perspective, it became appropriate to view 'representation' in terms of regions of state space, and 'computational rules' as attractors (Elman, 1995); the dynamics is supplied by Recurrent Neural Networks (RNN). Systems of this kind learn to recognize and generate languages after being trained on suitable examples. Surprisingly, it turned out that the induction of this ability occurs when small network parameter adjustments bring about a phase transition in the neural network's state space. Once in a certain state, machine states for correct recognitions scale with an exponent of 1.4 (Pollack, 1991). Considering the RNN as a dynamical system, it appears that its trajectories can locate regions in phase space which support fractal dynamics: putting it in a graphical way, the phase space would seem 'peppered' with regions for fractal dynamics (i.e. attractors), which can be reached by the trajectories of the complex system's dynamics. Numerous additional sources point to a close, though often not readily transparent relation between the dynamics of RNN and IFS: the principle is consistent with the observations of Pollack (1991) inasmuch as fractal sets provide a method for organizing recursive computation in a bounded state space (Tabor, 2000). Furthermore, context-free grammar computation by connectionist networks using fractal sets can generate spatial representations of symbolic sequences (Tino, 1999; Jeffrey ,1990) via IFS (Barnsley & Demko, 1985; Barnsley et al, 1989). A class of associative reinforcement learning algorithms was constructed by Bressloff and Stark (1992) as an extension of non-associative schemes in stochastic automata theory; within the IFS framework, it suggested a possibly fractal nature of the learning process. Tsuda and Kuroda (2004) recently elaborated this idea and developed a mathematical model of Cantor Coding for the formation of episodic memory in the hippocampus.

The intent is here merely to draw attention to a large segment of literature, of which the foregoing citations are but a small sample that implicates interrelations between state space dynamics of RNN and IFS in the processing of symbolic information. For clarification of this relationship, Kolen (1993, 1994) proposed that the RNN's state dynamics itself is an IFS, as a paradigmatic case of the synergism of fractal and complex system dynamics. Levy & Pollack (2001) obtained supportive evidence in that every point in the hidden layer of the RNN is either itself part of the fractal attractor of the IFS, or has an orbit that "ends on" the attractor in a finite number of steps. Two tantalizing questions arise: one, wherein does the 'computational' power of a Fractal System lie? How does the self-similar structures of fractals unpack layers of 'information" for guiding actions across many scales still eludes our comprehension. And, second, what exactly is the nature of that apparent synergism between complex system and fractal dynamics ? (see Section 5.2).

### 3.2 Motor Behavior and Allometric Control processes

During quiet standing, the human body sways in a seemingly erratic fashion. Collins and De Luca (1994) determined that the pattern of this postural sway is exhibits intrinsic correlations which can be modeled as a system of bounded correlated random walks. This result suggested to the authors that the postural control system incorporates both open and closed loop control mechanisms. The statistics of temporal patterns in spontaneous motor activity of laboratory rodents can be replicated by the stochastic mechanism of Davidsen and Schuster (2002, see Section 4) which generates power law distributions and 1 1/f power spectrum over several decades (Anteneodo and Chialvo, 2009). The presence of long-time correlations in the stride-interval time of normal humans suggested to Hausdorff et al (1995) that the activity of walking may be a self-similar fractal. Acknowledging the complexity of locomotor activity, the authors referred to its prerequisite of coordinating inputs from motor cortex, basal ganglia and cerebellum, as well as feedback from vestibular, visual and proprioceptive sources. In the Hausdorff et al. study, the gait cycle (synonymous with stride interval) was defined as the time between consecutive heel strikes of the same foot.

West & Griffin (1998) and Griffin et al (2000) took a different and novel approach to the analysis of gait patterns: using the time between consecutive maximal positive extensions of the same knee for measuring the stride interval, their data analysis was based on determining the long-time correlation properties of the stride interval time series. The Relative Dispersion (given as the ratio of standard deviation to the mean) for different levels of data aggregation captures the inter-relatedness of the data across multiple time scales. The method is described in detail by Bassingwhite et al, 1994) and is designed to answer whether the correlations are self-similar upon scaling (i.e.: identical between groups of neighbors at different time scales). The result of the data analysis was that the fluctuations of the gait cycle were self-similar with a fractal dimension of 1.25. In addition, stride-interval time series itself was in this study a random fractal, consistent with the data of Hausdorff et al.

The importance of the West-Griffin results lies in showing that the correlation in their data was an inverse power law of a form similar to the allometric scaling laws found in many areas of Biology : typically, allometry establishes a relation between two properties of an organism. Historically, the idea is based on Huxley's (1931) definition of allometric growth, describing that the different growth rates of two parts of an organism are proportional to one another. In the West-Griffin studies, the allometric principle is reflected in the constancy of the Relative Dispersion over the length of the stride interval time series which, in the present case, has the non-integer fractal dimension of 1.25. Their data raise the issue of systematic control of variability, which is generic of complex physiological processes. Unlike the familiar homeostatic control that regulates system variables by negative feedback, an allometric control system is conceived as regulating variability of a process involving multiple interactions among sensors and effectors with intricate feedback arrangements, each with its own characteristic set of frequencies and time scales. Their functions are reflected by the allometric relation which captures the process's long term memory with power law correlations, and by the power law distributions of the system variable (West, 1999 b).

The significance of this principle is documented by West (1999 a, 2006) for the numerous physiological processes which are identified as fractal, on the basis of their time series behavior. Notable examples of fractal Physiology are heart rate, bronchial air ways and body temperature variability, and integrated neural control networks. In these situations, the regulatory mechanisms constitute coupled cascades of feedback loops in systems far from equilibrium. Therapeutic interventions, commonly based on the homeostatic principle which assumes the significant system variable to be normally distributed fails to take the regulatory complexity into account and may be counterproductive. Instead, Allometric (fractional) Control based on Fractional Calculus (Podlubny, 1999) provides the appropriate approach. Applications of Fractional Calculus to modeling the interdependence and organization of complex system, such as for instance the vestibulo-oculomotor system, are illustrated by Magin (2004).

Changing walking speed, using metronomically controlled walking, or aging and pathological conditions introduce stress conditions to the neural control system which requires expanding the theoretical framework. Based on the notion of a stochastic model of human gait dynamics (Ashkenazy et al, 2002), West and Scafetta (2003) tested the model of a neural pattern generator on the data set obtained by Hausdorff et al, 1995) which they showed to exhibit slightly multifractal fluctuations. Metronome timing breaks the long-time correlations of the natural pace and generates a large fractal variability of the gait regime. The two essential features of the model required for capturing the phenomenology of the data set were that the dynamics of the system unfolds on an attractor in phase space, and that the natural frequency of the attractor is replaced by a random walk over a restricted set of frequencies which leads to the multifractal output for the dynamical model (Scafetta et al, 2009).

4. Processes that generate power law distributions

Antedating the modern theory of stochastic processes, Yule (1925) proposed a model of speciation to explain the highly skewed distributions of abundances of biological genera. Thirty years later, Simon (1955) derived several related stochastic processes from relatively general probability assumptions that lead to Yule-type distributions. Their characteristic properties distinguish them from the negative binomial and Fisher's logarithmic series. Leaving open the possibility of still other generative mechanisms for power law distributions, Simon suggests that the frequency of occurrence of this empirical distribution should not come as surprise. The preferential attachment scheme for network growth (Barabasi and Albert 1999) has stimulated the recent interest in the Yule-Simon approach in as much as Bornholdt and Ebel (2001) could show that they are closely related. The important step of introducing the notion of aging of network nodes was taken by Dorogovtsev and Mendes (2000): the probability of being linked to a newly added node is taken to be proportional to its current connectivity weighted by a power law function of its age. This motivated Cattuto et al (2006) to propose a modified Yule-Simon process that takes the full history of the system into account, applying a hyperbolic memory kernel.

Simon's conclusion that power law distributions can be derived from relatively general assumption seems to be born out by the number of mathematical models that have been proposed. A shot noise process, reviewed by Milotti (2002) is an example, as is the Reversible Markov Chain Models (Erland and Greenwood, 2007), and the Clustering Poisson Point Process (Grueneis,2001), the latter already introduced in Section 1.2. The simple stochastic mechanism of Davidsen and Schuster (2002) generates pulse trains with power law distributions of pulse intervals, and 1/f power spectra over several decades at low frequencies with an exponent close to 1. Iterated function systems (IFS) are a unified approach for generating and classifying a broad class of fractals with self-similarty (Barnsley and Demko , 1985). The Chaos Game is a generalized form of this, designating a method for generating the attractor (fixed point) of any IFS. Other Recurrence Models (Kaulakys et al, 1998, 2006) derive from a more specific frame of reference insofar as they consider random walks in complex systems that display self-organization. As alternative, Ruseckas and Kaulaskys (2010) generate 1/f noise with nonlinear stochastic differential equations. Touboul and Destexhe (2009) followed a similar route when developing their case against power law scaling of neural avalanches (see Section 2.3). Physical systems whose observable properties exhibit values which randomly exceed certain critical values are candidates for applying Extreme Value Theory: the aim of the classical form of this theory is to quantify the properties of the extremes (large or small) occurring in random sequences of independent numbers. Extremal dynamics may be applied to generate objects with fractal structure (Miller et al, 1993); as Extremal Optimization, it successively eliminates undesirable components of suboptimal problem solutions (Boettcher and Percus, 2000).

The various approaches discussed in the foregoing can essentially be viewed as *ad hoc* (Milotti ,2002). In contrast, however, there are two types of conceptual anchors that ground power law relations explicitly in larger foundational contexts. For one of the conceptual roots, I turn to the theory of Random Walks and fractional difference equations. The continuum limit of simple random walks is diffusion and, correspondingly, expressed in the mathematics of differential equations. The simple random walk aggregates the random steps from a large number of identically distributed random variables with finite variance. However, an extensive range of investigations has made it abundantly clear that simple random walks with this statistics do not capture the richness of biological data, and for that matter other fields of investigation as well (for reviews see : West and Deehring, 1995; West 1999; Bassingthwaighte et al, 1994). A decisive step beyond simple random walks was the introduction of the concept of Continuous-Time-Random Walk (CTRW) by Montroll and Weiss (1965). Some forms of CTRW are fundamentally different from the classical diffusion model by drawing the timing of steps from waiting time distributions, or by taking steps of randomly varying length. This is for instance the case when the waiting time distribution does not possess a characteristic time scale (for instance, has a power law distribution) : in this situation, the mean square displacement and the distribution of transition rates become fractal. Processes corresponding to these and related random walk models are then referred to as fractal random walks, corresponding to anomalous diffusion which occupies an important place for studying physical processes such as transport in disordered media or non-exponential (anomalous) relaxation of, for instance, glassy media. Along these lines, Montroll and West (1979), Hughes et al (1982) and others examined a large repertoire of stochastic processes with unusual probability distributions for the displacement per step. For certain parameters, these walks have infinite spatial moments, generate fractal self-similar trajectories, have characteristic functions with nonanalytic behavior, and lead to an analog of RNG transformations. In the continuum limit, the fractal random walk leads to the Fractional Langevin Equation of motion describing trajectories, and their ensemble densities, in phase space (West, 2006). Such processes are viewed as fractional kinetics, and mathematically addressed in fractional calculus (Sokolov et al, 2002; Kleinz and Osler, 2000) and by Fractal Operators (West et al, 2003).

In an application to Neuroscience, Lundstrom et al (2008) showed that neocortical pyramidal neurons' firing rate is a fractional derivative of slowly varying stimulus parameters: neuronal fractional differentiation effectively results in adaptation with many time scales (see Section 5.2). Fractional order dynamics of brainstem vestibulo-oculomotor neurons was demonstrated by Anastasio (1994) who also suggested that simulation of fractional-order differentiators and integrators can be approximated by integer-order high- and low-pass filters, respectively. Thus, fractional dynamics may possibly be applicable to motor control systems, generally. This is also suggested by the stride-interval time series of human gait being a random fractal, indicating the role of long-time correlations in walking (West and Griffin, 1999; see Section 3.1). Mandelbrot and van Ness (1968) defined Fractional Brownian Motions as a family of Gaussian random functions, parametrized according to the interdependence of successive increments, with the parameter ranging from zero (Gaussian Fractional Random Walk) to infinite in Fractional Brownian motion : the latter to account for the empirical studies of random phenomena with interdependence of distant samples. The conceptual connections to scaling invariance and to the theory of renormalization(Section 2.3.3) are discussed by Quian (2003). Fractional reaction-diffusion in inhomogenous media stabilizes steady state solutions of Turing patterns (Henry and Wearne, 2000).

In 1987, Shlesinger et al. introduced the Levy walk as a random walk with nonlocal memory, coupling space and time in a scaling fashion. For the alpha-stable Levy Walks, the transition probability varies with the size of the step (Montroll and West (1987). Anomalous diffusion results from a Levy Flight which is a process where the time taken to complete a transition depends on the length of the step (West et al, 1997). West et al (1994) also identified dynamical generators of Levy Statistics . In an elegant step towards unifying various classes of random walks, Zumofen and Klafter (1933) applied the framework of CTRW's to derive Levy stable processes. The interesting properties of Levy processes include their satisfying a scaling law, self-similarity and possessing memory (Allegrini et al, 2002) . Levy (1954) also generalized the Central Limit Theorem to include those phenomena for which the second moment diverges. West and Deering (1995) and West (2006) assembled a large number of data obtained from various biological systems that satisfy Levy walk statistics. In a motor skill acquisition task, Cluff and Balasubramaniam (2009) report that probability distributions for changes of fingertip speed in pole balancing are Levy distributed. In vitro recorded spontaneous electrical activity of neuronal networks exhibits scale –invariant Levy distributions and long-range correlations (Segev et al, 2002). This is thought to enable different size networks to self-organize for adjusting their activities over many time scales. Among animal movement patterns associated with random search behavior , Levy walks outperform fractional Brownian motion (Reynolds, 2009), presumably evolved under section pressure (Bartumeus, 2007).

Physical process models to account for fractal heavy-tailed distributions of traffic pattern of (information) packages in LAN's (Local Area Networks) are based on renewal reward processes, originally applied to commodity pricing (Taquu and Levy,1986). Applied to network package traffic, the model takes into account the presence of long packet trains ("on periods", with packages arriving at regular intervals) and long inter-train pauses ("off periods"). The superposition of many such packet trains displays on large time scales the self-similar behavior LAN's if the "on-off" distribution has infinite variance (Willinger et al, 1995, Willinger, 2000).

The second conceptual framework was already introduced in Section 2.3.3: power law distributions are among the novelties that arise in the vicinity of or at the critical point of a continuous phase transition, including criticality of the self-organized kind. This should not come as surprise since scaling reflects long-time correlations in the underlying process, analogous to the comparable re-ordering process at critical phase transitions (Wilson, 1979): both cases address a class of phenomena where events at many scales make contributions of equal importance. Significantly, the comprehensive review on Fractal structures in nonlinear dynamics by Aguirre et al. (2009) begins with the sentence "Fractal structures appear naturally in nonlinear dynamics, in such a way that the two concepts are deeply related". Their review draws particular attention to numerous instances in nature where attractor basin boundaries in dissipative (open) systems display fractal behavior. Giesinger's (2001) comment is a propos: " at one point, Bak(1996) considered SOC as universal, with scaling as consequence; it appears, however, that the balance of evidence shifted the question: *why is there scale invariance in Nature ?* to the question: *is Nature critical ?* "

For constructing theories that deal with problems that have multiple scales, the renormalization group (RNG) offers a general method. In Physics, the most frequently studied situation is 'percolation transition' for which Newman (2005) offers a detailed account of the origin of power law scaling: the cumulative distribution of cluster sizes forms at the critical point a power law distribution. Percolation transition is a special case under the closely interconnected family of RNG and coarse graining that entails power law distributions as a source of natural fractals (see Section 1.4). Coarse graining allows one to determine whether the phenomenon under investigation has universality, apart from scaling: Universality implies that macroscopic properties of a system are independent of the system's particular microscopic configuration. The particular values determined for a given instantiation of the system are then not significant, apart from showing that the system scales: the theoretical foundations are extensively discussed by Essam (1980) and by Stauffer and Aharony (1991/1994).

For Neuroscience, Kozma et al (2005) illustrated the potential relevance of percolation for phase transitions in models of neural populations with mixed local and global interactions, and (Werner,2009 b,c) proposed Renormalization Group Transformation as a general principle to account for functional relations between levels of neural organization. Since fractals will in both situations naturally arise, it is pertinent to ask what their role could be. West et al (2008) and Allegrini et al (2006) attribute to them a complexity matching function which will be the subject of review and comments in the next section 3.1.

## Section 5: Fractals in Action.

Having reached the end of the largely phenomenological surveys of fractal scaling and associated manifestations of fractality at the conventionally distinguished levels of organization and function of nervous systems, it appears inescapable to recall the title of Barnsley's (1993 ) book ‘Fractals everywhere”. The apparent prevalence in the nervous system is matched by the numerous manifestation in physiological systems generally (West and Deering, 1995). Is the ubiquity a sign of triviality, or the result of a generic and fundamental principle of Nature? If that is the case in Biology at least, then Nature seems to adhere to it with remarkable conservatism as, for instance, the monographs of Dewey (1999) for Molecular Biophysics and of Seuront (2009) in Ecology attest.

As documented by Aguirre et al. (2009) , and is referenced in several sections of this review: the intimate relation between fractals and nonlinear dynamics in dissipative systems is apparent and well substantiated. Hence the seemingly disproportionate attention paid to phase transitions and criticality in Sections 2.2 and 2.3. However, Section 4 lists a large number of alternatives for generating fractals, some of which obviously qualifying as ‘natural’, as, for instance Levy flights. Thus, the burden of proof of attributing observed fractals to nonlinear dynamics lies on identifying the fractal boundaries or the critical phase transition that gave rise to them. This subset of natural fractal must then be viewed as manifestation of, and a consequence of, the particular dynamic regime from which they originated.

### 5.1: the Complexity Matching Effect (CME)

The issue under consideration is the communication among complex systems generating fractal signatures. The starting point is the evidence presented in Section 1.4 that the EEG time series can be identified as a (non-ergodic) non-Poisson renewal (NPR) process, reflecting strong deviation from exponential decay. A brief account of CME will suffice at this point since a comprehensive overview of the underlying principle of CME is available in West et al, (2008). CME is concerned with the conditions under which one complex network responds to a perturbation by a second complex network: Consider a NPR network with a power law index < 2 as measure of its complexity, and apply a random signal as perturbation: this is in essence comparable to the condition of aperiodic Statistic Resonance (Gammaitoni et al, 1998). Allegrini et al (2006a,b) then generalized the conditions by applying as perturbation another complex network which also satisfies the NPR condition with power index < 2. Under these conditions, it can be shown that the effect of the perturbation is maximal if the power law indices of the interacting systems are equal. The claim is that CME, as illustrated in the foregoing, applies to a large class of NPRs such as, for instance, return times for random walks, either in regular lattices or in complex networks.

### 5.2: Linking actions across many scales ?

Even if Nature’s conservative adherence to fractals is a valid argument in support of their functional significance, we are still in the dark as to what that function may be. Emphasizing the feature of self-similarity, we can turn to types of functions which could benefit from stacking extended ranges of space and/or time scales into one compact format; moreover, there is no privileged time scale in power law dynamics. Sensory adaptation as a change over time in the responsiveness of the sensory system to a constant stimulus is a situation of this kind (Wark et al., 2007). Adaptation with power law dependence and multiple time scales has been demonstrated in nervous systems under many different conditions. Examples come from such diverse sources as electrosensory afferent nerve fibers in weakly electric fish (Xu et al,1996), spider mechanoreceptor neurons (French and Torkkeli, 2008). The auditory sensory memory which is thought to encode stimuli on multiple time scales (Ulanovsky et al., 2004). Fairhall et al (2001, b) direct attention to the speed with which the dynamics of a neural code is optimized, even when the statistical properties of the stimuli themselves evolve dynamically over a wide range of timescales, from tens of milliseconds to minutes. The source of 1/f fluctuations in human sensorimotor coordination tasks is presumably attributable to the multiple time scale activities of neural centers (Ding et al, 2002). In visual psychophysics, adaptation to contrast follows power law dynamics (Rose and Lowe, 1982) as does the tilt aftereffect (Greenlee and Mangussen, 1987). For fractional order dynamics in adaptation , see: Lundstrom,( 2008) and Anastasio (2004) in Section 4.

This fragmentary compilation of diverse observations is intended illustrate the propensity of various neural structures to respond swiftly to a range of temporal and/or spatial parameters: the image of a set of strings resonating to specific frequencies comes to mind. Could self-similar fractal structures, having stored a repertoire of responses, each specific for a particular range of temporal and/or spatial stimulus parameters, fulfil this function ? On formal grounds, Thorson and Biederman-Thorson (1974) attributed sensory adaptation to distributed relaxation processes, based on nonuniformities of local “efficacy” in the transduction process at peripheral receptors. Might this “local nonuniformity” be the expression of a self-similar functionality of receptors ?

Generically, models of adaptation integrate the response of a system and feed the integrated signal back to curtail that response. The type of adaptation is determined by the properties of the integrator. Applying this principle, Drew and Abbott (2006) examined a form of power law integration with the result that the suppressive effect of repeated stimuli on successive responses are accumulated with power law decline. In distinction from other forms of integration (e.g.: exponential), power law integration has the notable feature of scale-invariance and, in their simulations, replicated the published data of Xu et al (1996), cited in the previous paragraph. On the same principle, these investigators also implemented power law adaption within a standard spiking neuron model, except for approximating the time dependence of the power-law adaptation integral for computational efficiency by a series exponentials. This principle has been successfully applied by Hausdorff and Peng ( 1996) and is the basis of Anderson’s (2001) assertion of the power law being an emergent function : it amounts to linking the noninteracting exponential processes to a cascade which reproduces the power law forgetting in power law integration. The adaptation obtained of the integrate-and-fire model with adaptation current obtained by a cascade of exponential processes matched perfectly that obtained by injecting adaptation currents to the model neurons.

The implications of this successful approach are considerable: the Dell and Abbott results suggest that power law adaptation can be instantiated by a cascade of a large number of processes with ordinary exponential dynamics , covering a wide range of time constants. The cascaded model design lets the temporal stimulus dynamics set the appropriate adaptation dynamics, in virtue of the numerous exponential processes with different time scales. They present a telling argument in support of the biological significance of this mechanism of adaptation: natural stimuli vary unpredictably over a wide of time scales; instead keeping the recovery time after excitation constant, power-law adaptation allows the temporal statistics of the stimuli themselves to determine the dynamics of adaptation. The work of Toib et al ( 1998) and Gilboa et al ( 2005) referred to in Section 2.2.1 are additional examples of 'multiscale computing' involving fractals.

Are we prepared to envision a general principle of self-organizing control structures for multiscale behavior, extracting the statistics of an unpredictable environment by way of power law integration ? The papers cited in the next paragraph speak to this question.

Fusi et al (2005) applied the principle of power-law forgetting of adaptation to a cascade model for regulating the plasticity of synapses as the basis of stored memories. Note that memory strength is here represented as synaptic plasticity, not as synaptic strength, as is commonly the case. Each synapse model has two levels of synaptic strength, weak and strong. Associated with each strength is a cascade of states (in one of their models, five). The cascades introduce a range of probabilities for transition between weak and strong levels of synaptic plasticity. A complicated heuristics is built into the model design that enables combinations of states of plasticity, for instance: states with low probability are paired up with labile states, and conversely, etc. The important point is that a high level of memory storage with long retention times significantly outperforms other model designs. In a model of learning visuomotor associations that are reversed unpredictably from time to time, synaptic modification occurring on multiple time scales along the same principles can be the basis for flexible behavior; the model predictions were validated with experimental data. (Fusi et al, 2007).

It appears then, that the principle of shifting between different scales on demand manifests itself in many different forms, and on different levels of neural organization. Under natural conditions, are the required exponential functions for power law integration models of self-similar structures ?

## 6. Summary and final thoughts

The assembly of largely phenomenological data was presented to support the claim that fractal processes and properties occur at many and diverse levels of neural organizations and performance, and are functionally relevant. Several issues that must not glossed over lightly needed discussion in several places : e.g. fractality as such is not an obligatory indicator of SOC ; whether non-conservative systems may be limited to a state of 'quasi-criticality instead of being candidates for full criticality (Section 2.2); and what the origin of natural fractals may be (Section 4). However, the close relation of fractals with critical phase transitions is beyond dispute. Despite the evidence secured by Plenz and Chialvo in specially targeted experiments, critical dynamics is still by some investigators called in question at the mesoscopic level, but the evidence for its importance and essential role at the macroscopic domain is uncontroversial and solid.

Among the virtues of brain criticality discussed in Section 2.4.3., there is one that has attracted attention for the longest time: it is the ability for rapid changes of state. Phase transition in critical systems provide an universal mechanism for rapid switching between different cooperative neuron collectives. Less attention received the fact, well established in Physics, that the system presents at the critical transition qualitatively new properties, requiring new macroscopic descriptors. The important feature of the organization following the phase transition is to form new objects with distinct properties: a new Ontology. In Physics, transitions between ferro- and paramagnetism are the prototypical example. We need to ask: What, if any, is the manifestation of ontological novelty in brain phase transition ? Customarily, one tends to think in terms of integration and differentiation, brought about by the change in correlation patterns of the system components, with nearby clusters functioning in synchrony. However, this way of looking at the nature of the state change fails to meet the requirement for qualitative (ontological) novelty, as the analogy with Physics would require. In the paradigmatic cases in Physics, new aggregations are being formed on phase transition which display novel properties at a macroscopic level of description. What can we assume to happen in brain on phase transition ? The obvious answer is: partitioning into neural assemblies with fractal properties. This is of course merely analogous to the fractal patterns formed at phase transition in physical systems. But here is an essential difference between the systems studied in Physics and the brain: Take the ferro-paramagnetic phase transition again as example, it consist merely in a change of spin orientation of the elementary components. But in the case of brain, the elementary components are reactive neurons which have the potential of entering into aggregations with functional interactions. Model studies of Arcangelis and Herrman (2010) with self-organizing neuronal networks show that avalanches formed at phase transition can learn complex rules on the basis of a collective process. Under appropriate conditions, the learning dynamics is universal inasmuch as even complex rules can be acquired. Recalling Section 3.1, the relation of the state dynamics of these neural networks to Integrated Function Systems and fractal attractors would constitute a notable case of the synergism of fractal and complex system dynamics .

Re-visiting the work of Beggs and Plenz (Section 2.3) and putting it bluntly: phase transition endows avalanches with novel capacities that are not analytically predictable from the original state of the system. Generalizing from this, it is evident that pursuing these and related directions in targeted studies of the aftermath of brain phase transitions is imperative for seeking to gain a full appreciation of the functional novelties it may create. Whether such studies will show a way of bridging the ultimate barrier of the epistemic cut (Pattee, 2001) that separates the domain of integrable Physics from the domain of symbolic structures remains to be seen: but this issue is, I submit, at the core of the apparent object-subject (brain-mind) duality.

The limited sample of recent publication on network models cited in Section 2.4.1 is but a fraction of the large number of combinations and permutations of network design parameters that are conceivable: on the one hand, ranging from the of small-world to scale free works with various degree distributions, on other hand each type being hierarchical, modular with and without hubs, and fractal. Which one from among this collection is closest to the natural brain networks? Suggestions abound, but tantalizing questions remain still open to further inquiry. An architecture that does justice to self-similarity under renormalization appears attractive as it might unite the dynamics of network topology in one common mechanism, as I suggest in that Section.

Commenting on the wealth of existing data on anatomical and functional cortical networks organization may seem like "carrying vocals to Newcastle", notably in view of the comprehensive review prepared by Bullmore and Sporns (2009). Nevertheless, a few general considerations may be of some relevance. It is a commonplace observation that Nature loves Hierarchies: in the classical "Architecture of Complexity", Simon(1962) has given plausible reasons for this apparent love affair. A few years later, in a discussion of the organization of complex systems (Simon, 1973), observed that the term "hierarchy" has taken a somewhat generalized meaning, divorced from its original denotation in human organizations of a vertical authority structure. He there contended that in application to complex systems , "Hierarchy" has come to denote a set of Chinese Boxes of a particular kind, usually consisting of their recursion. But then, these Chinese Box hierarchies come in variants of one form or another of ordering, e.g. in the partial ordering of trees. There is then the possibility for a variety (and sometimes ambiguous) senses in which current investigators report their finding in anatomical, fMRI and model studies when speaking of hierarchical order.

The network and graph-theoretical community has developed its own criteria: modularity has been introduced a measure of a networks decomposability, for instance into community structures ( Newman and Girvan, 2004), possibly at different hierarchical levels, allowing one to "zoom in and out" for finding communities on different levels; this was discussed in detail by Meunier et al, (l.c.). Mucha et al (2009) seek to address this same objective in a new framework that encompasses coupling of individual networks via links connecting nodes of one with nodes of another network at multiple scales, thus allowing for some nesting which escapes the Newman-Girvan method, but is accessible by the approach of Sales-Pardo et al (2007). Keeping these issues in mind is, I suggest, of importance in the definitive evaluation of many of the reports of fMRI data on cortical functional organization: while describing modularity, they may not be able to capture nesting among modules, due to methodological limitations. Yet, it is the nesting of the kind that Simon refers to as "Russian doll" ( see also the Russian Matryoshka dolls of Agnati ,l.c.), where the pressing question of self-similarity arises: It has some plausibility in view of Sporns 's (2006) suggestive evidence for cortical connection patterns of fractal and self-similar nature. In different contexts, self-similar community structures were noted by Guimera et al (2003) in a network of human interactions, and nesting of theta- and beta gamma oscillations was noted by Gireesh and Plenz (2008) in neuronal avalanches of developing cortical layers 2/3.

The foregoing discussion was merely intended to draw attention to some of the problems involved in attempts to characterize dynamic hierarchies of a system that consists of multiple levels of organization, having dynamics within and between the entities described at each of the different levels (Lenaerts et al., 2005). Hierarchical nesting would presumambly qualify brain dynamics as a dynamic hierarchy. This is consonant with the recent findings of Honey et al (2009) of indirect functional connectivity, not accounted for by structural connections, and the variability of resting-state functional connectivity across scanning sessions and model runs. As regards Criticality, it would then be imperative to examine the properties of phase transitions in dynamically hierarchic systems. In one recent study of a kinetic Ising model, it turned out that the universality class is consistent with the two-dimensional equilibrium Ising model (Rikvold et al., 1999).

Sadly, despite the resourceful and imaginative studies undertaken by many investigators during the past decade, we are still some distance away from understanding the structural and functional basis of cortical dynamics; let alone the cortical-subcortical dynamics that, according to the Global Workspace Hypothesis (Baars,1988) is now generally considered decisive for cognitive functions.

Notwithstanding this disappointing state of affairs, one more speculations may perhaps be permitted: recall the amazing discovery from Section 5.2 : there, power-law forgetting of adaptation and synaptic modification on multiple time scales were discussed as principles for flexible behavior. Recall also the evidence from Section 2.2.1 for multiscale computing as basis cooperative fractal channel kinetics. The essential principle in these cases is the capacity of having available the responsiveness, on demand, to a large range of unpredictable external contingencies of a certain category, varying in one parameter: e.g. the capacity for selecting the level of adaptation appropriate to a wide range of temporal patterns and stimulus intensity. In the cases referred to, it amounted to scale shifting as required, involving fractality and self-similarity of control structures. If Sporns's evidence for cortical fractality and self-similarity holds up, I can now re-visit the claim I made in the Introduction: "is there a natural capacity for unpacking interactions between different levels of a fractal object or process, responsive to circumstances and conditions, which eludes us entirely? If it existed, fractals would surely be a most extraordinary design principle for operational economy in complex systems."

To which I now add: if this natural capacity embodies principles of power law integration and cascading, similar to those identified for adaptation and fractal kinetics, we may come to comprehend at least some aspects of the extraordinary functionality of the cortex.